\documentclass{aims} % Use the aims.cls file to compile your paper
\usepackage{amsmath}
\usepackage{paralist}
\usepackage[misc]{ifsym}
\usepackage{booktabs}
\usepackage{epsfig} %For pictures: screened artwork should be set up with an 85 or 100 line screen
\usepackage{epstopdf} %This is to transfer .eps figure to .pdf figure; please compile your article using PDFLaTex or PDFTeXify.
\usepackage[colorlinks=true]{hyperref}
\usepackage{multirow}
\hypersetup{urlcolor=blue, citecolor=red}
\allowdisplaybreaks

\def\currentvolume{X}
 \def\currentissue{X}
  \def\currentyear{200X}
   \def\currentmonth{XX}
    \def\ppages{X-XX}
     \def\DOI{10.3934/xx.xxxxxxx}

\usepackage{siunitx}
\usepackage{color}
\usepackage{colortbl}
\usepackage{bm}
\theoremstyle{definition}

\newcommand{\di}[1]{\,\mathrm{d}#1}
\newcommand{\Vdelta}{\boldsymbol{\delta}}
\newcommand{\conductivity}{{\rm Ohm}^{-1}}
\newcommand{\impedance}{{\rm Ohm}}
\usepackage{tikz}
\usetikzlibrary{patterns,matrix}
\usetikzlibrary{shapes,backgrounds}

\providecommand{\E}{{\mathbb E}}

\providecommand{\R}{{\mathbb R}}

\providecommand{\argmin}{\operatorname*{argmin}}				%argument yielding inf
\providecommand{\argmax}{\operatorname*{argmax}} 				%argument yielding max
\providecommand{\diag}{\operatorname{diag}}						%Diagonal

\providecommand{\Jacobian}{\mathbf{J}_{\sigma_\text{ref}}}
\providecommand{\Uref}{\mathbf{U}_\text{ref}}

\providecommand{\DeltaSigmaHat}{
\widehat{\delta \sigma}}

\providecommand{\sigmapix}{\boldsymbol{\sigma}}

\providecommand{\bepsilon}{\boldsymbol{\epsilon}}
\providecommand{\BSigma}{\boldsymbol{\Sigma}}
\providecommand{\FullyC}{\textsf{FC U-Net}}
\providecommand{\PostP}{\textsf{Post-Processing}}
\providecommand{\CondD}{\textsf{Conditional-Diffusion}}
\providecommand{\KTC}{\textsf{KTC2023}} % Maths macros

\usepackage{lineno}
\title[Data-driven approaches for electrical impedance tomography]
{Data-driven approaches for electrical impedance tomography image segmentation from partial boundary data} % Only the first word and proper nouns should be capitalized.

\author[Denker, Kereta, Singh, Freudenberg, Kluth, Maass and Arridge]{}

 % already filled in 

\thanks{$^*$Corresponding author: Alexander Denker}

\begin{document}
\maketitle

% Enter the first author's name and email address; email addresses are required for each author.
% Use footnote notations to indicate address and affiliations with commas between numbers if more than one address applies; see below for examples.
\centerline{\scshape
Alexander Denker$^{{\href{mailto:adenker@uni-bremen.de}{\textrm{\Letter}}}*1}$,
\v{Z}eljko Kereta$^{2}$,
Imraj Singh$^{2}$,   
}

\medskip

\centerline{\scshape Tom Freudenberg$^{1}$,
Tobias Kluth$^{1}$,
Peter Maass$^{1}$,
Simon Arridge$^{2}$}

\medskip

{\footnotesize
% Enter the full affiliation and country name:
% Do not insert commas or periods at the end of lines.
 \centerline{$^1$Center of Industrial Mathematics, University of Bremen}
} % Do not forget to end {\footnotesize with the sign }

\medskip

{\footnotesize
 % Enter the full affiliation and country name:
 \centerline{$^2$Department of Computer Science,  University College London}
}

\bigskip

% The name of the handling editor will be entered by AIMS production staff.
% "Communicated by Handling Editor" is not needed for special issue.
 \centerline{(Communicated by Handling Editor)}

%%%%%%%%%%%%%%%%%%%%%%%%%%%%%%%%%%%%%%%%%%%%%%%%%%%%%%%
%             5. ABSTRACT
%%%%%%%%%%%%%%%%%%%%%%%%%%%%%%%%%%%%%%%%%%%%%%%%%%%%%%%

\begin{abstract}
Electrical impedance tomography (EIT) plays a crucial role in non-invasive imaging, with both medical and industrial applications. In this paper, we present three data-driven reconstruction methods for EIT imaging. These three approaches were originally submitted to the Kuopio tomography challenge 2023 (KTC2023). First, we introduce a post-processing approach, which achieved first place at KTC2023. Further, we present a fully learned and a conditional diffusion approach. All three methods are based on a similar neural network as a backbone and were trained using a synthetically generated data set, providing with an opportunity for a fair comparison of these different data-driven reconstruction methods.
%This is the abstract of your article. It should not exceed \textbf{200} words and needs to be concise and factual. State the purpose of the research, the principal results, and conclusion.
\end{abstract}

%%%%%%%%%%%%%%%%%%%%%%%%%%%%%%%%%%%%%%%%%%%%%%%%%%%%%%
%                   6. BODY
%%%%%%%%%%%%%%%%%%%%%%%%%%%%%%%%%%%%%%%%%%%%%%%%%%%%%%

% This will be removed later
%\setcounter{section}{-1}
%\section{About the notation}
%We have three different representations of the conductivity distribution:
%\begin{itemize}
%    \item We use $\sigma$ as a function $L^\infty(\Omega) \to \R$ in Eqn.~\eqref{eq:maxwelleqn}
%    \item We use the same $\sigma$ as coefficients of the finite element basis, i.e., $\sum_{i=1}^M \sigma_i \chi_i$ with $\sigma = (\sigma_1, \dots, \sigma_M)$
%    \item We use $\sigmapix$ on the $d = 256 \times 256$ pixel grid, i.e., as an image 
%\end{itemize}

%Also, we have 
%\begin{itemize}
%    \item $M$ dimension of piecewise constant function space 
%    \item $N$ dimension of piecewise linear function space 
%    \item $K$ number of injection patterns 
%    \item $L$ number of electrodes 
%    \item $n$ number of training samples,$n_k$ number of training samples per level $k$
%    \item $d = 256 \times 256$ pixels 
%    \item $C (=3)$ classes (water, resistive, conductive)
%\end{itemize}
%Also, what are the correct units for the conductivity values?

% Only the first word and proper nouns of section titles should be capitalized.
% The title of section 1:
\section{Introduction}

Electrical impedance tomography (EIT) is an imaging modality that uses electrical measurements taken on the boundary of an object that are used to recover electrical properties of its interior. In this paper we consider the reconstruction of conductivity, for which a series of currents are applied through electrodes attached to the object's surface.
%\Simon{I think you don't mention permittivity again after this, which would require time harmomic measurements $u(\omega)$. In order not to over complicate the paper you could say something like "... in order to recover electrical properties of its interior. In this paper we consider only the reconstruction of conductivity, for which a series of currents..."} To reconstruct the conductivity, a series of currents are applied through electrodes attached to the object's surface.
The electrodes measure the resulting voltages, which are used to produce an image of the conductivity. EIT has numerous applications for example in medical diagnostics \cite{borsic2009vivo} or in non-destructive testing \cite{karhunen2010electrical}.

There are several mathematical models for the physics of the EIT measurement process. 
Let $\Omega$ be the domain, $\partial\Omega$ its boundary and $\cup_{l=1}^L e_l \subset \partial\Omega$ the set of $L\in\mathbb{N}$ electrodes attached to the boundary.  
The electric potential $u \in H_1(\Omega)$ is derived from Maxwell's equations and is governed by 
\begin{subequations}\label{eq:maxwelleqn}
\begin{align}
    \label{eq:maxwell1}
    -\nabla\cdot(\sigma \nabla u) = 0 \quad \text{in } \Omega,
\end{align}
where $\sigma \in L^\infty(\Omega)$ is the conductivity distribution.
%Applying the electrodes on the boundary $\partial\Omega$ produces an electric flux, which is assumed to be fully transferred to the boundary. This can be described by 
%\begin{equation}\label{eq:currents}
%    \sigma \frac{\partial u}{\partial \nu}=g \text{ on } \partial \Omega,
%\end{equation}
%where $g:\partial\Omega\rightarrow \R$ is the recorded voltage, and $\partial u/\partial \nu$ is the outward normal derivative.
%However, the continuum model is ill-suited for practical considerations as we only know the voltages at the electrodes.
%The complete electrode model (CEM)~\cite{somersalo1992existence} constructs a realistic mathematical framework by explicitly taking into account the effect the electrodes have on the measurement physics.
The complete electrode model (CEM)~\cite{somersalo1992existence} describes a realistic formulation of boundary conditions when a current is applied to the electrodes.
First, the boundary is decomposed into two components: the electrodes $e_l$ (identified with the part of the boundary they are attached to) and the remaining space between the electrodes, $\partial\Omega\setminus\cup_{l=1}^L e_l$.
Second, the electrical conduction between the electrode and the corresponding part of the boundary is accounted for.
For a given current injection pattern $I \in \R^L$, the resulting model can be written as
\begin{align}\label{eq:CEM}
\begin{cases}
    u + z_l \sigma \frac{\partial u}{\partial \nu} = U_l, &\text{on } e_l, \text{ for } l=1,\dots,L, \\ 
    \sigma \frac{\partial u}{\partial \nu} = 0, &\text{on } \partial \Omega \setminus \cup_{l=1}^L e_l, \\
    \int_{e_l} \sigma \frac{\partial u}{\partial \nu} ds = I_l, &\text{on } e_l, l=1,\ldots,L,
\end{cases}
\end{align}
where $z \in \R^L$ are the contact impedances, quantifying the effect of the resistive layer formed at the contact point of electrodes and the boundary, and $U \in \R^L$ is the voltage at the electrodes. CEM includes conservation of charge and a mean-free current constraint for the potentials, i.e., 
\begin{align}\label{eq:conservation_of_charge}
    \sum_{l=1}^L I_l = 0 \text{ and } \sum_{l=1}^L U_l = 0,
\end{align}
\end{subequations}
respectively. 
We denote the voltage as $\mathsf{U}=(U_1,\ldots, U_L)^\top$ and the current pattern as $\mathsf{I}=(I_1,\ldots, I_L)^\top$. 
Equations \eqref{eq:maxwell1} to \eqref{eq:CEM} describe a single current injection pattern. In practice, several injection patterns are applied and corresponding electrode measurements are obtained.
We denote the voltages and charges for the $k$-th injection pattern by $\mathsf{U}^{(k)}$ and $\mathsf{I}^{(k)}$, where  $k=1,\ldots,K$. 
By $\mathbf{U}=(\mathsf{U}^{(1)},\ldots, \mathsf{U}^{(K)})$ and $\mathbf{I}=(\mathsf{I}^{(1)},\ldots, \mathsf{I}^{(K)})$ we denote stacked $\R^{KL}$ vectors of voltages and charges at all electrodes and for all current patterns. Let further $\mathbf{F}(\sigma) = (\mathsf{F}^{(1)}(\sigma),\ldots, \mathsf{F}^{(K)}(\sigma))^\top$ be the corresponding forward operator, applied to conductivity $\sigma$, for all electrodes and current patterns.
The resulting non-linear inverse problem can be written as 
\begin{align}
    \label{eq:nonlinear_ip}
    \mathbf{F}(\sigma)\mathbf I = \mathbf U.
\end{align}
where the goal is to reconstruct $\sigma$ given electrode measurements $\mathbf{U}$.

\subsection{KTC 2023 Challenge}\label{subsec:ktc_challenge_description}
We outline three methods submitted to the Kuopio Tomography Challenge 2023 (\KTC) \cite{rasanen2023kuopio}.
The goal of \KTC~was to reconstruct segmentation maps of resistive and conductive inclusions from partial boundary measurements. %to use EIT measurements of given 2D targets to provide a segmentation of the reconstructed image into background (water), and resistive and conductive inclusions.
The measurements were acquired from a plastic circular tank with $32$ equispaced stainless electrodes attached to the boundary.
%Electrodes were enumerated counterclockwise, with the first electrode corresponding to the $12$ o'clock position. 
%Each electrode covers a $5.625$ degree range, with a further $5.625^\circ$ degrees between neighbouring electrodes.
All $32$ electrodes were used to collect the measurements and $16$ electrodes (the odd numbered ones) were used for current injection patterns.
For each current injection pattern voltages are taken between adjacent electrodes, resulting in $31$ measurements per injection pattern.
Five types of injection patterns are considered. An illustration of the measurement tank, electrodes, and injection patterns can be found in Figure~\ref{fig:eit_illustration}.

%\begin{description}
%    \item[Adjacent] Current between electrodes $e_1\rightarrow e_3$, $e_3\rightarrow e_5$, \ldots, $e_{31}\rightarrow e_1$;
%    \item[All against $1$] Current between electrodes $e_3\rightarrow e_1, \ldots, e_{31}\rightarrow e_1$;
%    \item[All against $9$] Current between electrodes $e_1\rightarrow e_9, \ldots, e_{31}\rightarrow e_9$;
%    \item[All against $17$] Current between electrodes $e_1\rightarrow e_{17}, \ldots, e_{31}\rightarrow e_{17}$;
%    \item[All against $25$] Current between electrodes $e_1\rightarrow e_{25}, \ldots, e_{31}\rightarrow e_{25}$.
%\end{description}

{
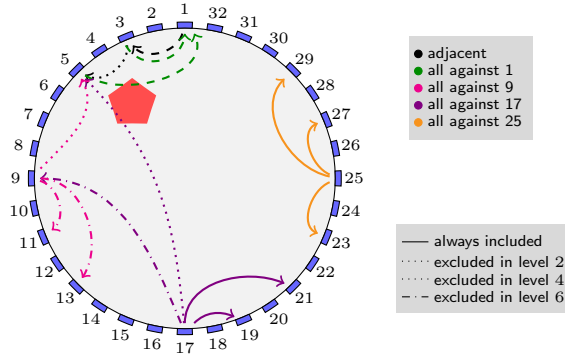
\begin{figure}\label{fig:eit_illustration}
\centering
\begin{tikzpicture}
    \def\n{32} % Number of nodes
    \def\radius{2cm} % Radius of the circle
    \def\biggerradius{2.08cm}
    \def\smallradius{0.05cm} % Width of each node
    \draw[fill=gray!10] (0,0) circle (\radius);
    \node[star, star points=5, star point ratio=1.25, fill=red!70, inner sep=0.19cm] at (-0.7,1.0) {};
    \foreach \i in {1,...,\n}
    {
         \filldraw[fill=blue!60,opacity=1.0] ({-2.8125+(\i-1)*11.25}:\radius) arc ({-2.8125+(\i-1)*11.25}:{2.8125+(\i-1)*11.25}:\radius) -- ({2.8125+(\i-1)*11.25}:\biggerradius) arc ({2.8125+(\i-1)*11.25}:{-2.8125+(\i-1)*11.25}:\biggerradius) -- cycle;
        \node at ({360/\n * (\i-1)+90}:\radius+0.25cm) {\tiny{\i}};
        \node (\i) at ({360/\n * (\i-1)+90}:{\radius+0.05cm}) {};
    }
        \draw[dashed, ->, thick] (1) to [out={75+180},in={127.5+180}] (3);
        \draw[dotted, ->, thick] (3) to [out={97.5+180},in={150+180}] (5);

        \draw[dashed, ->, green!55!black, thick] plot [smooth, tension=1.5] coordinates { (3.south) (-0.328408, 1.56783) (1.south)};
        \draw[dashed, ->, green!55!black, thick] (5) to [out={180-212.5},in={115+180}] (1.south east);

        \draw[->, yellow!50!red, thick] (25) to [out={180+22.5},in={180-22.5}] (23);
        \draw[->, yellow!50!red, thick] (25) to [out={180-22.5},in={180+22.5}] (27);
        \draw[->, yellow!50!red, thick] (25) to [out={180-11.25},in={180+45.}] (29);

        \draw[->, blue!50!red, thick] (17) to [out={180-135},in={180-22.5}] (19);
        \draw[<-, blue!50!red, thick] (21) to [out={180-22.5},in={180+270}] (17);
        \draw[dashdotted, ->, blue!50!red, thick] (17) to [out={180-90+22.5},in={0}] (9);
        \draw[dotted, ->, blue!50!red, thick] (17) to [out={180-90+5},in={-45}] (5.south east);

        \draw[dotted, <-, magenta, thick] (5) to [out={180+165-45},in={180+180+45}] (9);
        \draw[dashdotted, ->, magenta, thick] (9) to [out={180+165},in={180+217.5}] (11);
        \draw[dashdotted, <-, magenta, thick] (13) to [out={180-123.5},in={180+180}] (9);
        \matrix[fill=gray!30, row sep=-1mm, inner sep=2pt, column 1/.style={anchor=base west}, column 2/.style={anchor=west}] at (3.8,1.2) {
          \node [draw, shape = circle, fill = black, minimum size = 0.1cm, inner sep=0pt] (){}; & \node[black,font=\tiny] {\textsf{adjacent}}; \\
          \node [draw=green!55!black, shape = circle, fill = green!55!black, minimum size = 0.1cm, inner sep=0pt] (){}; & \node[black,font=\tiny] {\textsf{all against 1}}; \\
          \node [draw=magenta, shape = circle, fill = magenta, minimum size = 0.1cm, inner sep=0pt] (){}; & \node[black,font=\tiny] {\textsf{all against 9}}; \\
          \node [draw=blue!50!red, shape = circle, fill = blue!50!red, minimum size = 0.1cm, inner sep=0pt] (){}; & \node[black,font=\tiny] {\textsf{all against 17}}; \\
          \node [draw=yellow!50!red, shape = circle, fill = yellow!50!red, minimum size = 0.1cm, inner sep=0pt] (){}; & \node[black,font=\tiny] {\textsf{all against 25}}; \\
        };
        \matrix[fill=gray!30, row sep=-0.5mm, inner sep=2pt, column 1/.style={anchor=base west}, column 2/.style={anchor=west}] at (4.0,-1.2) {
            \draw (0,0) -- (1em,0); & \node[black,font=\tiny] {\textsf{always included}}; \\
            \draw[dotted] (0,0) -- (1em,0); & \node[black,font=\tiny] {\textsf{excluded in level $2$}}; \\
            \draw[dotted] (0,0) -- (1em,0); & \node[black,font=\tiny] {\textsf{excluded in level $4$}}; \\
            \draw[dashdotted] (0,0) -- (1em,0); & \node[black,font=\tiny] {\textsf{excluded in level $6$}}; \\
        };
\end{tikzpicture}
\caption{An illustration of the EIT measurement tank ($\Omega$), the electrodes $e_l$, with a sample of the injection patterns. In black we show the \emph{adjacent} injections; in green \emph{all against} $e_1$; in pink \emph{all against} $e_9$; in magenta \emph{all against} $e_{17}$; in orange \emph{all against} $e_{25}$. Dashed injection are removed in the $2^{\text{nd}}$ challenge level; dotted ones in the $4^{\text{th}}$; dash dotted in the $6^{\text{th}}$.}
\end{figure}
}

%Adjacent injection patterns consist of $16$ injections, and the remaining patterns consist of $15$ injections. Thus, there are $76$ overall injection patterns, each with $31$ measurements, giving a total of $2356$ measurements when all the data is used.

The challenge was divided into $7$ difficulty levels, where the first level included data from all $32$ electrodes. In subsequent levels, pairs of electrodes were successively removed.
Consequently, the number of measurements and the number of applied current injection patterns decrease with each level. Each level two more electrodes are removed in successive order, i.e., level $2$ removes electrodes $1,2$ and in the final level $7$ electrodes $1$ to $12$ are removed. This means, that all measurements from the upper left boundary are removed, cf. Figure~\ref{fig:eit_illustration}.  % and the measurements corresponding to injections that include any of the excluded electrodes. 
This made the reconstruction of the conductivity map and segmentation map increasingly ill-posed for higher levels. %Moreover, reconstruction and segmentation is more difficult for objects that have been immersed close to the part of the boundary where the electrodes have been removed.

%\addtolength{\tabcolsep}{-2pt}    
%\begin{table}[]\small
%\centering
%\begin{tabular}{l|ccccccc}
%Level    & 1 & 2 & 3 & 4 & 5 & 6 & 7 \\ \hline
%\multirow{2}{1cm}{Removed electrodes}   & N/A &$e_1, e_2$ & $e_1,\ldots, e_4$ & $e_1,\ldots, e_6$ & $e_1,\ldots, e_8$ & $e_1,\ldots, e_{10}$ & $e_1,\ldots, e_{12}$  \\
%\\
%measurements    & 2356 & 1624 & 1404 & 1200 & 1012 & 630 & 513       
%\end{tabular}
%\caption{Description of challenge levels with the removed electrodes and the number of measurements for each level \TODO{table needs improving: alignment is a bit wrong}}
%\end{table}
%\addtolength{\tabcolsep}{2pt}    

\subsection{Our Contribution}
\label{sec:contribution}
We propose three data-driven reconstruction methods to tackle the reconstruction of segmentation maps from partial EIT measurements:
\begin{itemize}
\item \FullyC: a fully-learned approach that reconstructs directly from measurements, see Section~\ref{sec:FCUNet} .
\item \PostP: an approach that reconstructs from an initial reconstruction method, see Section~\ref{sec:post_unet}.
\item \CondD: a conditional diffusion approach that aims to directly model the posterior given initial reconstructions, see Section~\ref{sec:cond_score}.
\end{itemize}

Both \FullyC~and \PostP~are learned reconstruction approaches, whereas \CondD~is an approach based on conditional generative modelling. The three proposed methods achieved the three highest scores at \KTC, with \PostP~performing the best overall. Additionally, the three approaches use a similar U-Net architecture with a comparable number of parameters and are trained using a dataset of generated phantoms and simulated measurements.

\subsection{Related Work}
\label{sec:relatedwork}
Deep learning post-processing and fully learned reconstruction are two well-known data-driven approaches for medical image reconstruction~\cite{arridge2019solving}. Both of these frameworks have been applied to EIT image reconstruction. Our \FullyC~ follows the model proposed by Chen et al.~\cite{chen2020deep}. However, in Section~\ref{sec:FCUNet} we propose a novel two-step training method for this fully learned model. Post-processing methods have been applied to EIT, e.g., by Martin et al.~\cite{martin2017post}. We extend this post-processing framework to deal with the different levels, corresponding to partial EIT measurements with increasing severity, of the KTC2023. To the best of our knowledge, our submission is the first application of a conditional diffusion model to real-world EIT data. Recently, Wang et al.~\cite{wang2024comparative} propose the use of an unconditional diffusion model and make use of the sampling framework proposed by~Chung et al.~\cite{chung2022come} to enable conditional sampling. However, they only evaluate their approach on simulated data with two or four circular inclusions, whereas we evaluate our approach on real measurements of complex objects.

%% I think that this should be in the earlier section. We should perhaps provide a general overview of the approach first somewhere, and only then go into the 
% Instead of using the provided forward solver of the organizers we used a custom implementation. 
% The reasons are, that we wanted an in-depth understanding of the solver to customise it and apply different initial reconstruction methods (see Section \ref{sec:initial_reconstruction}) as well as optimize the computation time to create the dataset. 

\section{Linearised Reconstruction}
EIT reconstruction deals with the recovery of the conductivity $\sigma$ from a set of measurements of electrode measurements~$\mathbf{U}$. A common technique is to linearise the non-linear forward operator $\mathbf{F}$ around a homogeneous background conductivity $\sigma_\text{ref}$ and reconstruct a perturbation $\delta \sigma$ to this background~\cite{cheney1990noser,dobson1994image,kaipio2000statistical}. The linearised forward operator is given as 
\begin{align}
    \Tilde{\mathbf{F}}(\sigma_\text{ref} + \delta \sigma; \sigma_\text{ref}) \mathbf{I} := \mathbf{F}(\sigma_\text{ref})\mathbf{I} + \Jacobian \delta \sigma,
\end{align}
where $\Jacobian := \nabla \mathbf{F}(\sigma_\text{ref})$ is the Jacobian evaluated at the background conductivity. We further assume access to measurements $\Uref$ of the empty water tank, such that $\mathbf{F}(\sigma_\text{ref})\mathbf{I} = \Uref$.
%\Simon{I don't know if you want to mention it, but this usually involves some "calibration" step e.g. $\mathbf{F}(\sigma) \rightarrow \frac{\Uref}{{\mathbf{F}}(\sigma_\text{ref})} \mathbf{F}(\sigma) $; maybe no need especially if space is short. }
We then define the measurement perturbation as $\mathbf{\delta U} := \mathbf{U} - \Uref$. The corresponding linearised problem is then to determine $\delta \sigma$ from 
\begin{align}
    \label{eq:linearised_ip}
    \Jacobian \delta \sigma = \mathbf{\delta U}.
\end{align}
%Above is discretised using piecewise constant finite element expansion for perturbation is represented by $\delta \sigma \in \R^M$.  , this is explained in Sections \ref{sec:forward_operator} and \ref{sec:jacobian}. The  coefficients of a .
%\zeljko{rephrase/move}
The perturbation $\delta \sigma \in \R^M$ is discretised by $M$ coefficients of a piecewise constant finite element expansion. %and assume that $\Jacobian$ is discretised accordingly. 
The finite element approximations of the Jacobian, and forward operator, are explained Sections \ref{sec:forward_operator} and \ref{sec:jacobian}. 

To solve Eqn.~\eqref{eq:linearised_ip} we use the framework of variational regularisation as 
\begin{equation}
    \label{eq:var_probl}
    \DeltaSigmaHat := \argmin_{\delta \sigma}\frac{1}{2} \|\Jacobian \delta \sigma -  \mathbf{\delta U} \|^2_{\BSigma^{-1}} + \alpha \mathcal{J} (\delta \sigma),
\end{equation}
where we assume a Gaussian noise model $\mathbf{\delta U}\sim\mathcal{N}(\mathbf{0}, \BSigma)$, and $\mathcal{J}: \R^M \to \R_{\ge }$ is a regulariser.
We consider Tikhonov-type regularisers of the form $\mathcal{J}(\delta \sigma) = \frac{1}{2} \| \mathbf{L} \delta \sigma \|_2^2$. For this choice of a regulariser we can recover the solution to Eqn.~\eqref{eq:var_probl} as 
\begin{align}
    \label{eq:lin_rec_one_step}
    \DeltaSigmaHat = (\Jacobian^\top \BSigma^{-1} \Jacobian + \mathbf{L}^\top \mathbf{L})^{-1} \Jacobian^\top\BSigma^{-1} \mathbf{\delta U}.
\end{align}
The matrix $(\Jacobian^\top \BSigma^{-1} \Jacobian + \mathbf{L}^\top \mathbf{L})^{-1}$ can be computed offline, leading to a computationally cheap reconstruction method necessary for training the post-processing and conditional diffusion networks. In the following sections, we will discuss the implementation of the forward operator, the computation of the Jacobian $\Jacobian$ and the choice of regulariser $\mathcal{J}$. 

\subsection{Forward Operator} \label{sec:forward_operator} 
The EIT forward operator $\mathbf{F}$ defining CEM is non-linear. Evaluating $\mathbf{F}$ for a given a conductivity $\sigma$ requires solving the differential equations in \eqref{eq:CEM}. 
We approximate the solution by applying the finite element method to the  weak formulation of the CEM, see e.g. \cite{hyvonen2004complete}.
To incorporate the conservation of charge we introduce a Lagrange multiplier $\lambda \in \R$. The weak formulation of the CEM then reads: find $(u, \mathsf{U}, \lambda) \in H^1(\Omega) \times \R^L \times \R$ such that
% \begin{align}\label{eq:CEM_weak_formulation}
%     \int_{\Omega} \sigma \nabla u \nabla v \di{x}
%     +
%     \sum_{l=1}^L \frac{1}{z_l} \int_{e_l} &(U_l - u)(V_l - v) \di{s} = \sum_{l=1}^L I_l V_l,
% \end{align}
% for all $(v, V) \in H^1(\Omega) \times \R^L$. To incorporate the conservation of charge (\ref{eq:conservation_of_charge}) we introduce a Lagrange multiplier $\lambda$ and modify the weak formulation: Find a tuple $(u, U, \lambda) \in H^1(\Omega) \times \R^L \times \R$ such that 
\begin{align}\label{eq:CEM_weak_formulation}
    \int_{\Omega} \sigma \nabla u \cdot \nabla v \di{x}
    +
    \sum_{l=1}^L \frac{1}{z_l} \int_{e_l} &(u - U_l)(v - V_l) \di{s} 
    + \sum_{l=1}^L \left(\lambda V_l + \nu U_l\right)
    = \sum_{l=1}^L I_l V_l,
\end{align}
for all $(v, \mathsf{V}, \nu) \in H^1(\Omega) \times \R^L \times \R$, with $\mathsf{V}=(V_1, \dots, V_L)^\top$ and $\mathsf{U}=(U_1, \dots, U_L)^\top$. 
% The advantage of the Lagrange multiplier is that the problem stays symmetric. 

To numerically approximate the forward model we use the Galerkin approximation to the CEM, see e.g. \cite{kaipio2000statistical}
We give a short summary below. 
%For the finite element approximation we use a uniform mesh on the circular water tank $\Omega$, see Section~\ref{sec:practical_initReco} for further details. 
We represent the electric potential $u$ using piecewise linear basis functions $\{ \phi_i \}_{i=1}^N$, spanning a finite dimensional subspace~$V_N$ of $H^1(\Omega)$. 
The conductivity $\sigma$ is represented using piecewise constant basis elements $\{\chi_i\}_{i=1}^M$, where each $\chi_i$ is the indicator function of exactly one simplex in the mesh. 
To simplify the notation we identify $u$ and $\sigma$ with the coefficients in their respective basis expansions. That is, $u\approx \sum_{i=1}^N u_i\phi_i\cong(u_1,\ldots,u_N)^\top$ and 
analogously for $\sigma \approx \sum_{j=1}^M \sigma_j \chi_j \cong(\sigma_1, \ldots, \sigma_N)^\top$.

Applying the above Galerkin approximation to \eqref{eq:CEM_weak_formulation} results in the linear system
%Representing $u$ in the finite element basis, meaning $u = \sum_{i=1}^N u_i \phi_i$, and testing with each basis function $\phi_i$ we obtain the linear system
\begin{equation}\label{eq:linear_system_cem}
    \begin{pmatrix}
        A(\sigma) + B & C   & \mathbf{0} \\
        C^\top   & D   & \mathbf{1} \\
        \mathbf{0}^\top     & \mathbf{1}^\top & 0
    \end{pmatrix}
    \begin{pmatrix}
        u \\
        \mathsf{U} \\
        \lambda
    \end{pmatrix}
    = 
    \begin{pmatrix}
        \mathbf{0} \\
        \mathsf{I} \\
        0
    \end{pmatrix}, 
\end{equation}
with block matrices
\begin{alignat*}{2}
    A_{ij} &= \int_\Omega \sigma \nabla \phi_i \cdot  \nabla \phi_j \di{x} \quad \quad &&\text{for } i,j=1,\dots, N
    \\
    B_{ij} &= \sum_{l=1}^L \frac{1}{z_l} \int_{e_l} \phi_i \phi_j \di{s} &&\text{for } i,j=1,\dots, N
    \\
    C_{ij} &= \frac{1}{z_j} \int_{e_j} \phi_i \di{s} &&\text{for } i=1,\dots,N \text{ and } j=1,\dots, L
    \\
    D_{ii} &= \frac{1}{z_i} \int_{e_i} 1 \di{s} &&\text{for } i=1,\dots,L
\end{alignat*}
where 
% $u=(u_1, \dots, u_N)^\top \in \R^N$ \Simon{maybe you don't need this as it was writtend just above eq.(\ref{eq:linear_system_cem}) } and 
$\mathsf{U}=(U_1,\ldots, U_L)^\top\in\R^L$.

There are two properties of the linear system~\eqref{eq:linear_system_cem} that can be used to reduce the computational effort. First, for a fixed conductivity $\sigma$, the CEM is linear with respect to the injection patterns. This enables reusing intermediate steps of the procedure for solving Eqn.~\eqref{eq:linear_system_cem}, e.g., the LU factorisation of the system matrix which is used to compute the numerical solution $(u, \mathsf{U})$ for all current patterns $\mathsf{I}^{(k)}$ under consideration. Second, only the block matrix $A(\sigma)$ depends on $\sigma$, and needs to be recomputed. All other block matrices can be computed offline. 

The resulting discrete forward operator is implemented with the finite element software FEniCSx \cite{BarattaEtal2023}, and is available online\footnote{\url{https://github.com/alexdenker/eit_fenicsx}}.

%\subsection{Estimating background conductivity}
%The background conductivity and contact impedance were estimated based on the provided reference measurements of the empty water tank. In particular, the homogeneous conductivity $\sigma_\text{ref}$ was estimated using least squares fitting \cite{vilhunen2002simultaneous}. The recovery of the background conductivity and the contact impedance can be formulated as the following optimisation problem
%\begin{align}
%    (\sigma_\text{ref}, (z_l)_l) = \argmin_{\sigma, z} \sum_{k=1}^K \| U_{I^k}(\sigma, z) - U_\text{ref}^k \|_2^2
%\end{align}
%for current patterns $I^k$ with $k=1,\dots,K$. Due to convergence issues with the electrical impedance values, we opted to use a constant electrical impedance of~$z=\num{1e-6}$ for all electrodes. The background conductivity was estimated as~$\sigma_\text{ref} = 0.745$ \TODO{(units?)}, which was then used for the rest of our experiments. 

\subsection{Jacobian}
\label{sec:jacobian}
% In addition to the forward operator, we also need access to the Jacobian $\Jacobian$. 
We compute the Jacobian $\Jacobian$ using the discrete functions spaces for electric potential and conductivity. 
Alternative computational strategies using pixel grids or with the adjoint differentiation are demonstrated in \cite{dobson1994image, kaipio2000statistical}. 

Given $K$ injection patterns and $L$ electrodes, the Jacobian $\Jacobian$ is an ${LK\times M}$ matrix. 
However, it is perhaps more intuitive to view the Jacobian as an $L\times K\times M$ tensor. Using \cite[Appendix]{GEHRE20122126}, the Jacobian can be expressed as 
\begin{align}
    \label{eq:construct_jacobian}
    (\Jacobian)_{\cdot, k, j} = \mathsf{W}_{k,j}, \text{ for }  k=1,\dots,K, ~ j=1, \dots, M, 
\end{align}
where $(w_{k,j}, \mathsf{W}_{k,j}) \in V_N \times \R^L$ is the solution to 
\begin{align}
\label{eq:jacobi_problem}
    \int_\Omega\!\! \sigma_\text{ref} \nabla w_{k,j}\!\cdot\! \nabla \phi_i \di{x} \!+\!\! \sum_{l=1}^L\! \tfrac{1}{z_l}\!\! \int_{e_l}\!\! (w_{k,j} \!-\! (\mathsf{W}_{k,j})_l)(\phi_i \!-\!  V_l) \di{s}\! =\! -\! \int_{\Omega}\!\! \chi_j \nabla u^k\!\cdot\!\nabla \phi_i \di{x}, 
\end{align}
with $\sum_{l=1}^L (\mathsf{W}_{k,j})_l = 0$, where $u^k$ is the potential corresponding to current pattern~$\mathsf{I}^k$. 
%To construct $\Jacobian$ we need to compute for $k=1, \dots K, j=1, \dots, M$ the discrete solution $(w_{k,j}, W_{k,j}) \in V_N \times \R^L$ that fulfills for all $i=1, \dots, N$ and $ V\in \R^L$
%\begin{equation}\label{eq:jacobi_problem}
%    \int_\Omega \sigma_\text{ref} \nabla w_{k,j} \cdot \nabla \phi_i \di{x} + \sum_{l=1}^L \tfrac{1}{z_l} \int_{e_l} (w_{k,j} - W_{k,j}) (\phi_i -  V) \di{s} = - \int_{\Omega} \chi_j \nabla u^k \cdot \nabla \phi_i \di{x}, 
%\end{equation}
%as well as $\sum_{l=1}^L (W_{k,j})_l = 0$.
%Here $u^k$ is the electrical potential given by Eqn. (\ref{eq:maxwelleqn}) corresponding to the current pattern $I^k$ and reference conductivity $\sigma_\text{ref}$. Then it holds $(\Jacobian)_{\cdot, k, j} = W_{k,j}$, for a derivation of this identity we point to \cite[Appendix]{GEHRE20122126}. 
Similarly to Eqn.~\eqref{eq:CEM_weak_formulation}, we introduce a Lagrange multiplier to deal with the constraints, leading to to the same system matrix as in Eqn.~\eqref{eq:linear_system_cem} but with a different right hand side,
\begin{equation}
    \label{eq:linear_system_jacobian}
    \begin{pmatrix}
        A(\sigma_\text{ref}) + B & C   & \mathbf{0} \\
        C^\top   & D   & \mathbf{1} \\
        \mathbf{0}^\top     & \mathbf{1}^\top & 0
    \end{pmatrix}
    \begin{pmatrix}
        w_{k,j} \\
        \mathsf{W}_{k,j} \\
        \lambda_{k,j}
    \end{pmatrix}
    = 
    \begin{pmatrix}
        f_{k,j} \\
        \mathbf{0} \\
        0
    \end{pmatrix},
\end{equation}
with
\begin{alignat}{2}
    (f_{k,j})_i = - \int_\Omega \chi_j \nabla u^k \cdot \nabla \phi_i \di{x} \quad \quad &&\text{for } i=1,\dots, N.
\end{alignat}
Using the identity in Eqn.~\eqref{eq:construct_jacobian}, $K\cdot M$  problems need to be solved to construct the Jacobian~$\Jacobian$. 
However, since the dimensionality of the right hand side in Eqn.~\eqref{eq:linear_system_jacobian}, i.e. the range of the forward operator, is at most $N$ it suffices to compute the solutions $(w_r, \mathsf{W}_r, \lambda_r) \in V_N \times \R^L \times \R$ of   
\begin{equation}
    \begin{pmatrix}
        A(\sigma_\text{ref}) + B & C   & \mathbf{0} \\
        C^\top   & D   & \mathbf{1} \\
        \mathbf{0}^\top     & \mathbf{1}^\top & 0
    \end{pmatrix}
    \begin{pmatrix}
        w_{r} \\
        \mathsf{W}_{r} \\
        \lambda_{r}
    \end{pmatrix}
    = 
    \begin{pmatrix}
        \Vdelta_r \\
        \mathbf{0} \\
        0
    \end{pmatrix}
\end{equation}
where  $\Vdelta_r=(\delta_{ir})_{i=1}^N\in \R^N$ is the $r$-th unit vector. Thus, we only need to solve $N$ linear systems\footnote{Moreover, we have $N<M$, i.e., the dimension of piecewise linear elements is lower than the dimension of piecewise constant elements.}, instead of $K \cdot M$. 
As $f_{k,j}$ can be represented as a linear combination of $\{\Vdelta_r\}_{r=1}^N$, we can recover the Jacobian as
\begin{equation}
    \label{eq:fast_jacobian}
    (\Jacobian)_{\cdot, k, j} = \mathsf{W}_{k,j} = \sum_{r=1}^N (f_{k,j})_r \mathsf{W}_r.  %\int_{\Omega} \chi_j \nabla u^k \cdot \nabla \phi_r \di{x}.
\end{equation}
Observe that the piecewise constant elements $\chi_j$ are non-zero on exactly one element of the mesh. Thus, only a few summands on the right hand side in Eqn. \eqref{eq:fast_jacobian} remain, further reducing the computational complexity. 

In the higher challenge levels, boundary electrodes are removed. This results both in fewer electrode measurements $\Tilde{L} < L$ and fewer injection patterns $\Tilde{K} < K$, i.e., the reduced Jacobian is of shape $\Tilde{L}\Tilde{K} \times M$. We can compute this reduced Jacobian, by removing the corresponding rows of the full Jacobian $\Jacobian$.

%To speed up the computation we note that Eqn. (\ref{eq:jacobi_problem}) is linear with respect to the right-hand side. Additionally, for the dimension of continuous piecewise linear finite elements, it usually holds that $N < M$. Therefore, instead of solving (\ref{eq:jacobi_problem}) we compute for $r=1,\dots, N$ the discrete solution $(w_{r}, W_{r}, \lambda_r) \in V_N \times \R^L \times \R$ of 
%\begin{equation}
%    \begin{pmatrix}
%        A(\sigma_\text{ref}) + B & C   & 0 \\
%        C^\top   & D   & E \\
%        0     & E^\top & 0
%    \end{pmatrix}
%    \begin{pmatrix}
%        w_{r} \\
%        W_{r} \\
%        \lambda_{r}
%    \end{pmatrix}
%    = 
%    \begin{pmatrix}
%        -e_r \\
%        0 \\
%        0
%    \end{pmatrix}
%\end{equation}
%with the $r$-th unit vector $e_r\in \R^N$. Since $f_{k,j}$ can be represented as a linear combination of $\{e_r\}_{r=1}^N$ and by the linearity of the problem it follows
%\begin{equation}
%    (\Jacobian)_{\cdot, k, j} = W_{k,j} = \sum_{r=1}^N W_r \int_{\Omega} \chi_j \nabla u^k \cdot \nabla \varphi_r \di{x}.
%\end{equation}
%Since $\chi_j$ is non-zero solely on one element of the mesh, only a few summands on the right-hand side remain and we can quickly compute the complete Jacobian. 

\subsection{Regularisation}\label{sec:initial_reconstruction}
\begin{figure}
    \centering
    \includegraphics[width=1.\textwidth]{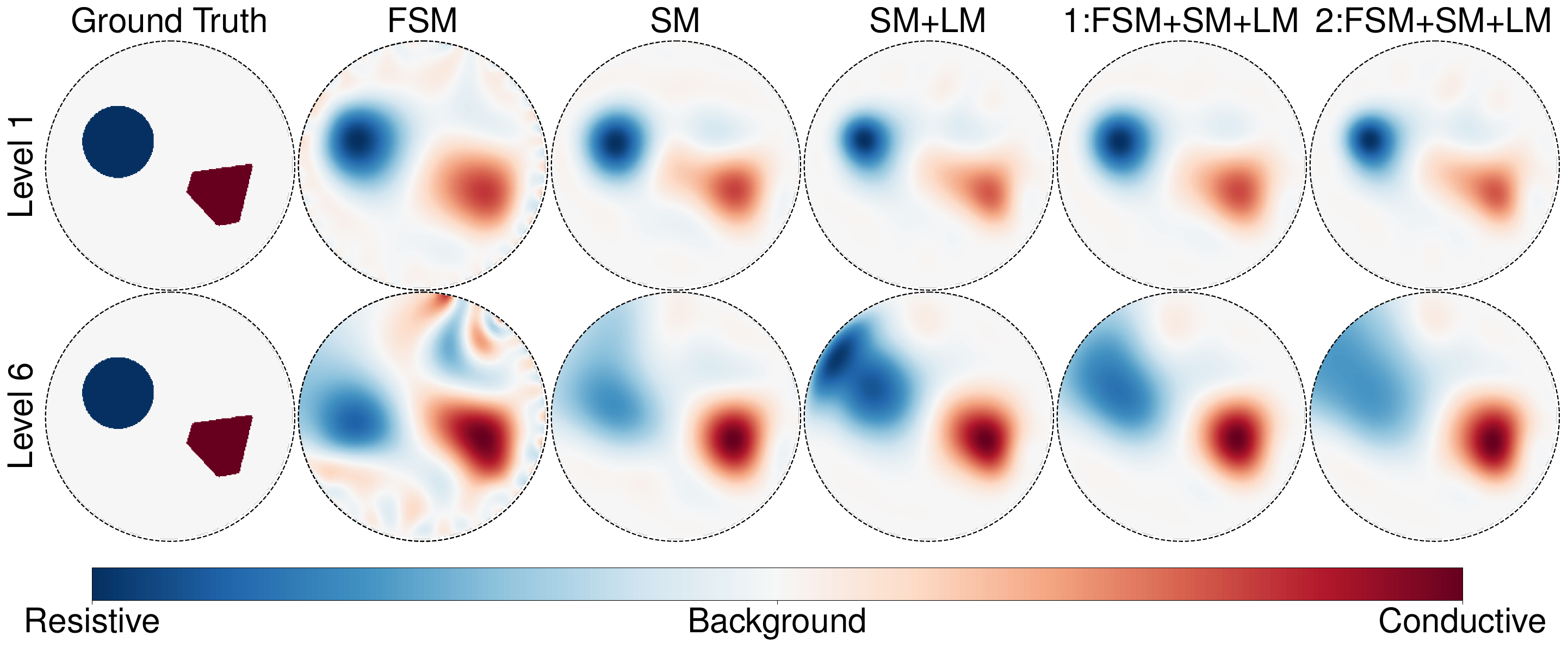}
    \caption{Example initial reconstructions on challenge levels 1 and 6. Level 6 was chosen as it best highlights differences in linearised reconstructions. We evaluate an independent FSM prior, independent SM prior, joint SM+LM prior and two joint priors FSM+SM+LM with different regularisation strengths. The chosen image is a sample of the validation data.}
    \label{fig:initial_reco}
\end{figure}

% Linearised reconstructions depend on the measurement perturbation~$\delta\mathbf{U}$, background conductivity~$\sigma_\text{ref}$, and the choice of the regulariser $\mathbf{L}$. The homogeneous background conductivity~$\sigma_\text{ref}$ is used to linearise the forward model and is estimated using least squares fitting based on the provided reference measurements~\cite{vilhunen2002simultaneous}. 
%% Commented out since it is repetitive
We consider Tikhonov-type regularisers $\mathcal{J}(\delta \sigma) =\frac{1}{2} \| \mathbf{L} \delta \sigma \|_2^2$. Note that for the reconstruction in Eqn.~\eqref{eq:lin_rec_one_step} we need access to $\mathbf{L}^\top \mathbf{L}$. Thus, we can define $\mathbf{P} := \mathbf{L}^\top \mathbf{L}$ instead of $\mathbf{L}$. We use three different regularisers:  
\begin{itemize}
    \item \textbf{First-order smoothness prior (FSM)}: %The FSM prior is constructed using 
    We define the mesh Laplacian $\mathbf{P}_\text{FSM} \in \R^{M \times M}$ with 
    \begin{align}
        (\mathbf{P}_\text{FSM})_{i,j} = \begin{cases}
            \text{deg}(i) & \text{ if } i=j \\ 
            -1& \text{ if } i \neq j \text{ and } i \text{ is adjacent to } j\\
            0 & \text{ else,}
        \end{cases} 
    \end{align}
    where $\text{deg}(i)$ is the number of neighbours of mesh element $i$. Matrix $\mathbf{P}_\text{FMS}$ can also be defined as $\mathbf{P}_\text{FSM} =\mathbf{L}_\text{FMS}^\top \mathbf{L}_\text{FMS}$, for $\mathbf{L}_\text{FMS}$ constructed as in \cite{borsic2009vivo}.
    %; each row $\mathbf{L}_i$ has two non-zero values, $1$ and $-1$ for the two adjacent mesh elements akin to a first-order approximation of the gradient. $M$ is the number of mesh elements and $N_e$ the number of adjacent mesh elements.
    \item \textbf{Smoothness prior (SM)}:  Smoothness distance matrix is constructed via $\mathbf{P}_\text{SM} := \mathbf{\Sigma}_\text{SM}^{-1}$ with $(\mathbf{\Sigma}_\text{SM})_{i,j} = a \exp(- \| x_i - x_j \|_2^2 / (2 b^2))$ where $x_i$ and $x_j$ are the coordinates of mesh elements $i$ and  $j$. We choose $a=0.025$ and $b= 0.4 \cdot 0.115$. This prior was used in the implementation provided by the organisers of the \KTC{}  \cite{rasanen2023kuopio}.
    \item \textbf{Levenberg–Marquardt regulariser (LM)} \cite{fletcher1971modified}: The LM regulariser is used in the NOSER framework \cite{cheney1990noser} and is defined as $\mathbf{P}_\text{LM} = \diag(\Jacobian^\top \BSigma^{-1} \Jacobian)$. Note that $\BSigma$ is the covariance matrix of the Gaussian noise model in Eqn.\eqref{eq:var_probl}.
\end{itemize}
In summary, the regularised solution to Eqn.~\eqref{eq:var_probl} is obtained by combining the three regularisers as 
\begin{equation}
    \label{eq:initial_reco}
    \begin{split}
            %\DeltaSigmaHa
            \mathbf{F}^\dagger(\delta \mathbf{U}) \!=\! (\Jacobian^\top \BSigma^{-1} \Jacobian \!+\! \alpha_\text{FSM} \mathbf{P}_\text{FSM} \!+\! \alpha_\text{SM} \mathbf{P}_\text{SM}   \!+\! \alpha_\text{LM} \mathbf{P}_\text{LM}%\diag(\Jacobian^\top \BSigma^{-1} \Jacobian) 
            )^{-1} \Jacobian^\top \BSigma^{-1} \mathbf{\delta U},
    \end{split}
\end{equation}
where $\alpha_\text{FSM}$, $\alpha_\text{SM}$, and $\alpha_\text{LM}$ are the regularisation strengths, and $\mathbf{F}^\dagger$ is the linearised reconstruction operator. 
The regularisation strengths are selected using a validation set of the four measurements provided by the organisers. 
The chosen regularisers promote different structures within the reconstructed images. Moreover, different regularisation choices produce clearly distinct reconstructions with corresponding artefacts, which is especially evident for  higher challenge levels. 
% No single regularisation strength is optimal for all measurement perturbations; and each regularizer promotes
For \PostP~ and \CondD~ approaches we use $5$ different regularisation choices, as shown in Figure~\ref{fig:initial_reco}.
The guiding idea is that combining the information from the various reconstructions will improve the performance of the trained convolutional neural network. 

The linearised reconstruction computed from Eqn.~\eqref{eq:initial_reco} resides on the piecewise constant mesh representation, whilst convolutional neural networks require inputs represented as a $256 \times 256$ pixel grid. Bilinear interpolation, denoted as $\mathcal{I}:\R^M \to \R^{256^2}$, was used to interpolate from mesh to image. We denote the resulting set of five interpolated linearised reconstructions as

%As convolutional neural networks require inputs to be images, an interpolation, denoted as $\mathcal{I}:\R^M \to \R^{256^2}$, from the piecewise constant mesh representation to the $256 \times 256$ pixel grid, is needed. We denote the resulting set of five interpolated linearised reconstructions as

\begin{equation}
    \boldsymbol{\DeltaSigmaHat} := \{ \mathcal{I} (\mathbf{F}^\dagger_j(\delta \mathbf{U}) )\}_{j=1}^{5}
\end{equation}
where the subscript denotes the $j$-th choice of regularisation strengths. In fact the choice of regularisation varies between challenge level for all five linearised reconstructions, i.e., a weaker regularisation is required for the full view setting at level $1$, meaning that a total of $35$ variations of regularisation strengths are defined. For clarity we omit an index for the challenge level.

\section{Deep Learning Approaches}
We submitted three deep learning approaches to the KTC2023. Two learned reconstructors, \FullyC~ and \PostP, and a generative approach, \CondD. All of our approaches share the same U-Net backbone\footnote{Accessible at \url{https://github.com/openai/guided-diffusion}} \cite{ronneberger2015u,dhariwal2021diffusion}. This network includes a conditioning mechanism allowing the level/timestep to more effectively influence the models output.

% architecture with adaptive group normalisation to incorporate level or timestep information. The network architecture only differs in the number of input channels for the first convolutional layer.
%\begin{itemize}
%    \item \FullyC: $1$ input channel 
%    \item \PostP: $5$ input channels, for the different initial reconstructions
%    \item \CondD: $6$ input channels, for the different initial reconstructions and the noisy image.
%\end{itemize}
% The \KTC~ thus offers us the opportunity to compare these different data-driven approaches, when implemented using the same network architecture. 
All models were trained on a simulated data set, which is described in Section~\ref{sec:dataset}. 
% In the challenge, the reconstructions were compared against segmentations on a $256 \times 256$ pixel grid and not against representations on a mesh. 
% To this end, 
Let $\sigmapix \in \R^d$ with $d=256^2$ denote the representation, discussed in Section~\ref{sec:initial_reconstruction}, of the reconstruction on the square pixel grid, where the pixels outside of the circular water tank are always treated as the background class. %\Simon{Do you actually consider a full pixel grid, or only those pixels inside the domain? if the latter then the dimension of $\sigmapix$ becomes a bit smaller}. \TODO{We actually consider the full pixel grid, as the reconstruction were judged based on the SSIM on the full $256 \times 256$ grid (however, we make sure that everything outside the water tank is always set to $0$, i.e., the background class}

\subsection{Learned Reconstructors}

The goal in learned reconstruction is to identify parameters $\hat{\theta}$ of a parametrised reconstruction operator $\mathcal{R}_\theta: \R^{KL} \to \R^d$, such that 
\begin{align}
    \mathcal{R}_{\hat{\theta}}(\delta \mathbf{U}) \approx \delta \sigmapix.
\end{align}
Given a paired data set $\{(\delta \mathbf{U}^{(i)}, \delta \sigmapix^{(i)}) \}_{i=1}^n$ of samples, we compute 
\begin{align}
    \hat{\theta} = \argmin_{\theta} \frac{1}{n} \sum_{i=1}^n \mathcal{L}(\delta \sigmapix^{(i)}, \mathcal{R}_\theta(\delta \mathbf{U}^{(i)})),
\end{align}
using a suitable loss function $\mathcal{L}:\R^d \times \R^d \to \R_{\ge 0}$. The mean-squared error loss function is commonly employed for reconstruction tasks \cite{ongie2020deep}. However, the goal in the challenge was not to reconstruct the conductivity distribution, but rather to provide a segmentation into water/background, resistive and
conductive inclusions. 
Therefore, we use categorical cross entropy (CCE) as a loss function. CCE is commonly used for image segmentation but has also been used for computed tomography segmentation \cite{arndt2023model}. 
Let $\mathcal{R}_{\hat{\theta}}$ denote the learned reconstructor. The model outputs logits, which are transformed to class probabilities by using a softmax function
\begin{align}
    \hat{p}_{i, c} := \text{Softmax}(\mathcal{R}_{\hat{\theta}}(\delta \mathbf{U})_{i, \cdot}) :=   \frac{\exp(\mathcal{R}_{\hat{\theta}}(\delta\mathbf{U})_{i, c})}{ \sum_{c' = 1}^C \exp(\mathcal{R}_{\hat{\theta}}(\delta\mathbf{U})_{i, c'})},
\end{align}
for all pixels $i=1, \dots, d$ and all classes $c=1, \dots, C$. Let further $p \in  \{0,1\}^{d \times C}$ be the one-hot encoding of the ground truth class. The CCE loss is defined as
\begin{align}
    \mathcal{L}_\text{CCE}(\hat{p}, p) = - \frac{1}{d} \sum_{i=1}^{d} \sum_{c=1}^C p_{i,c} \log(\hat{p}_{i,c}).
\end{align}
After training, the final segmentation is obtained by choosing the class with the highest probability, i.e., $\argmax_{c} \hat{p}_{i, c}$ at each pixel $i=1, \dots, d$.
%\begin{align}
%    \hat{\sigmapix}_i := \argmax_{c = 1, \dots, C} \hat{p}_{i, c}, \text{ at each pixel }i=1, \dots, d.
%\end{align}
The network is directly trained for segmentation, thus avoiding the need for an additional segmentation step. For both learned reconstruction methods, \FullyC~ and \PostP, we provide the challenge level as an additional input to the model and train a single model for all levels. 
They differ in the parametrisation of the reconstruction operator $\mathcal{R}_\theta$. Where the \FullyC~ implements a neural network directly acting on the measurements, the \PostP~ defines a two-step approach~\cite{ongie2020deep,schwab2019deep}.

\subsubsection{\FullyC}
\label{sec:FCUNet}
%First, we propose a fully learned neural network approach, called \FullyC, mapping directly from measurements to segmentations. 
The design of the \FullyC~ closely follows the work of Chen et al.~\cite{chen2020deep}. The model consists of two components: an initial learned transformation that maps the measurements to a pixel grid and a subsequent segmentation, implemented as a convolutional neural network. 
% \zeljko{this might be confusing, ie we don't need to }

\begin{figure}
    \centering
    \includegraphics[width=\textwidth]{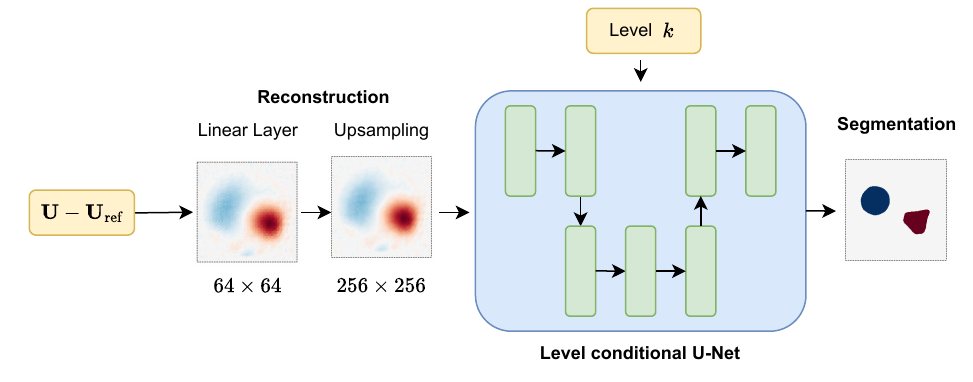}
    \caption{\FullyC~ network. We first use a linear layer to map the measurements to a $64 \times 64$ pixel grid, this is then bilinearly interpolated to the $256 \times 256$ grid. The network is trained to output class probabilities using categorical cross-entropy loss. The class probabilities are converted to segmentation maps by assigning the class with highest probability.}
    \label{fig:fcUNet}
\end{figure}

Instead of using the linear reconstruction method from Section \ref{sec:initial_reconstruction}, we will learn a linear mapping (represented as a single fully connected linear layer) that is applied to the measurements. However, learning a linear mapping from the measurements, with dimension~$KL~=~2356$, to the pixel grid, would require more than $150$M parameters and is computationally intractable. To reduce the number of parameters, we only learn a mapping to a $64 \times 64$ pixel grid and use a bilinear interpolation to the $256 \times 256$ pixel grid. The output of this initial transformation is used as an input to the second stage. Let $W \in \R^{64^2 \times KL}$ denote the initial linear layer and $\mathcal{S}: \R^{64^2} \to \R^{256^2}$ be the bilinear upsampling operator. The \FullyC~ is given by 
\begin{align}
    \mathcal{R}_\theta(\delta \mathbf{U}, k) := \Tilde{\mathcal{R}}_\theta(\mathcal{S}(W \delta \mathbf{U}), k) \quad k=1, \dots, 7,
\end{align}
with $k$ being the challenge level and $\Tilde{\mathcal{R}}_\theta$ is implemented as the attention U-Net \cite{dhariwal2021diffusion}. An overview of this approach is given in Figure~\ref{fig:fcUNet}. The missing measurements in~$\delta \mathbf{U}$ for the higher challenge level are filled with zeros.

To learn the linear map $W$, for the initial reconstruction, and the weights $\theta$ for the segmentation, we propose a novel two phase training process. 
In the first phase only the initial linear layer is trained using a mean-squared-error loss 
\begin{align}
\min_{W} \sum_{k=1}^7 \sum_{i=1}^{n_k} \| \mathcal{S}(W \delta \mathbf{U}^{(k,i)}) - \boldsymbol{\delta} \sigmapix^{(k,i)} \|_2^2. 
\end{align}
%where $n_k$ is the number of training examples per level $k=1,\dots,7$.
The aim of this phase is to provide a good initialisation of $W$. 
Afterwards, the full model is trained to provide a segmentation using the CCE loss 
\begin{align} \label{eq:cce}
    \min_{\theta, W} \sum_{k=1}^7 \sum_{i=1}^{n_k} \mathcal{L}_\text{CCE}(\hat{p}^{(k,i)}, p^{(k,i)}), 
\end{align}
where $\hat{p}^{(k,i)} = \text{Softmax}(\Tilde{\mathcal{R}}_\theta(\mathcal{S}(W \delta \mathbf{U}^{(k,i)}),k))$. In this joint optimisation of $\theta$ and $W$, we used a smaller learning rate for the linear layer $W$ than for $\theta$.

The dataset used for training the $\FullyC$ consisted only of random phantoms and simulated measurements. When evaluated on four challenge phantoms provided by the organisers, we noticed a deterioration in the final segmentation. To alleviate this generalisation problem, we added a finetuning phase, where the $\FullyC$ was trained for $1000$ optimisation steps on these $4$ challenge phantoms using a small learning rate of $\num{1e-6}$, for both $W$ and $\theta$.

\subsubsection{\PostP}
\label{sec:post_unet}  
\begin{figure}
    \centering
    \includegraphics[width=\textwidth]{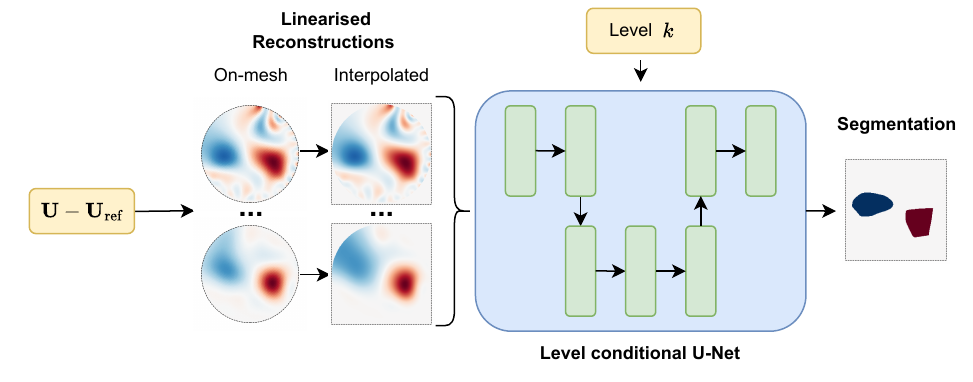}
    \caption{\PostP~ network. The five linearised reconstructions are interpolated to the pixel grid as described in Section~\ref{sec:initial_reconstruction}. The network is trained to output class probabilities using categorical cross-entropy loss. The class probabilities are converted to segmentation maps by assigning the class with highest probability.}
    \label{fig:postprocessingUNet}
\end{figure}

Learned post-processing was one of the first applications of deep learning to inverse problems \cite{arridge2019solving,ongie2020deep}. In this approach an initial reconstruction (computed from a classical reconstruction method) is used as an input to a convolutional neural network. More precisely, the reconstruction operator is parametrised as $\mathcal{R}_\theta(\delta \mathbf{U}) = \Tilde{\mathcal{R}}_\theta( \mathbf{F}^\dagger(\delta \mathbf{U}))$ where $\mathbf{F}^\dagger(\delta \mathbf{U})$ denotes the initial reconstruction. 
We adapt this approach in three ways.
First, a bilinear interpolation step is used to map the mesh values to an image for the convolutional neural networks. Second, five linearised reconstructions are used as initial reconstructions, cf. Section~\ref{sec:initial_reconstruction}. Last, the network is conditioned on the challenge level, as for the \FullyC, cf. Section~\ref{sec:FCUNet}. These adaptions result in the following formulation:
\begin{align}
    \mathcal{R}_\theta(\delta \mathbf{U}, k) = \Tilde{\mathcal{R}}_\theta(\boldsymbol{\DeltaSigmaHat}, k), \quad k=1, \dots, 7,
\end{align}
where $\boldsymbol{\DeltaSigmaHat}$ are the five interpolated linearised reconstructions and $k$ is the challenge level. An overview of the \PostP~is given in Figure~\ref{fig:postprocessingUNet}. The resulting network is trained for segmentation using the CCE loss function \ref{eq:cce} over all challenge levels and training samples, with the predicted class probability given by 
\begin{align}
    \hat{p}^{(k,i)} = \text{Softmax}(\Tilde{\mathcal{R}}_\theta( \boldsymbol{\DeltaSigmaHat}^{(k,i)} , k)), \text{ for } k=1,\dots,7, \text{ and } i=1,\ldots, n_k.
\end{align}

\subsection{Conditional Density Estimation}
From a statistical perspective of inverse problems, we are interested in recovering the posterior distribution $p^\text{post}(\sigmapix|\mathbf{U})$, i.e., the conditional distribution of conductivity $\sigmapix$ given the boundary measurements~$\mathbf{U}$~\cite{stuart2010inverse}. The goal in conditional density estimation is to approximate the true posterior $p(\sigmapix|\mathbf{U})$ with a conditional probabilistic model $p_\theta(\sigmapix | \mathbf{U})$ given a data set $\{ (\sigmapix^{(i)}, \mathbf{U}^{(i)}) \}, i=1, \dots, n$ with $(\sigmapix^{(i)}, \mathbf{U}^{(i)}) \sim p(\sigmapix, \mathbf{U})$. 
% The trained probabilistic model then allows us to evaluate different estimators. Here we work directly with the pixel representation $\sigmapix$.
%For our submission, we make use of the conditional mean (CM) \TODO{We dont really use the mean, do we?}
%\begin{align}
%    \sigma_\text{CM}(\mathbf{U)} = \E_{\sigma \sim p^\text{post}(\sigma|\mathbf{U})}[\sigma] \approx \frac{1}{K} \sum_{i=1}^K \sigma^k \quad \sigma^k \sim p_\theta(\sigma| \mathbf{U}),
%\end{align}
%which is approximated using samples from the conditional probabilistic model. 
In this work $p_\theta(\sigmapix | \mathbf{U})$ is modelled using denoising diffusion probabilistic models~(DDPM)~\cite{ho2020denoising,sohl2015deep}, which have shown promising results on many image generation tasks \cite{dhariwal2021diffusion}.

\subsubsection{Conditional diffusion models}
\label{sec:cond_score}
%Recently score-based~\cite{hyvarinen2005estimation} and denoising diffusion probabilistic models (DDPM)~\cite{ho2020denoising,sohl2015deep} have shown promising results in various image generation tasks~\cite{dhariwal2021diffusion}.
%Both approaches have been unified in the framework of score-based diffusion models~\cite{song2020score} using a stochastic differential equation to implement the diffusion process. 
Conditional variants of diffusion models were proposed for various inverse problems, including super-resolution \cite{saharia2022image}, time series imputation~\cite{tashiro2021csdi} and image inpainting~\cite{batzolis2021conditional}. Specifically, we build on ideas from \cite{batzolis2021conditional}. 

\begin{figure}
    \centering
    \includegraphics[width=\textwidth]{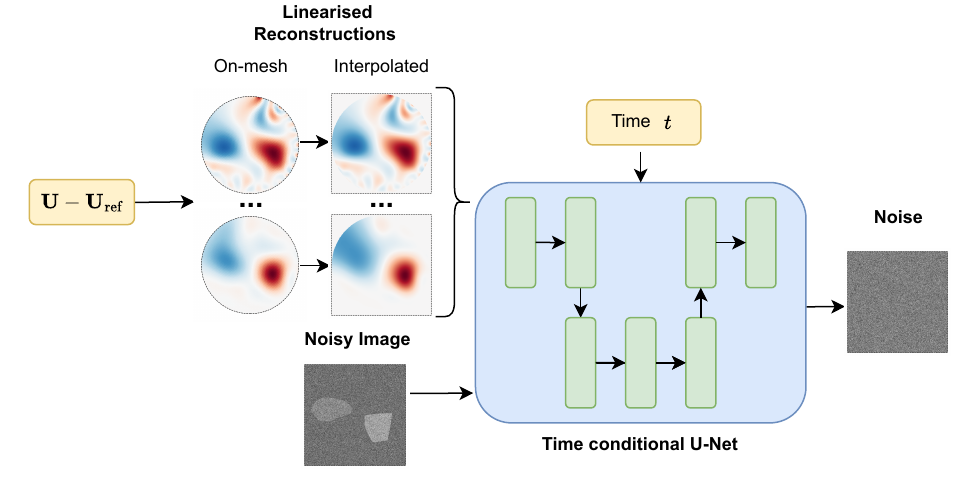}
    \caption{\CondD~ network. The five linearised reconstructions are interpolated to the pixel grid as described in Section~\ref{sec:initial_reconstruction}. The noisy image and linearised reconstructions are input into the network. Using the $\bepsilon$-matching loss function, the network is trained to estimate the noise. Through sampling the network a segmentation map is obtained. Multiple samples are drawn through pixel-wise majority voting the final segmentation map is obtained.}
    \label{fig:conddiffUNet}
\end{figure}

We make use of the discrete time formulation of diffusion models~\cite{ho2020denoising}. DDPMs define a forward diffusion process, given by a Markov chain, which gradually adds noise to the data over $T=1000$ timesteps.
\begin{align}
    \sigmapix_t = \sqrt{1 - \beta_t} \sigmapix_{t-1} + \sqrt{\beta_t} \bepsilon, \quad \bepsilon \sim \mathcal{N}(\mathbf{0},\mathbf{I}),
\end{align}
with variances $\beta_1 \le \dots \le \beta_T$. The variances are chosen so that the terminal distribution approaches a standard Gaussian,  $\sigmapix_T \sim \mathcal{N}(\mathbf{0},\mathbf{I})$. Given the noiseless sample $\sigmapix_0$, the noisy sample at time $t$ can be directly obtained as 
\begin{align}
    \label{eq:forward_onestep}
    \sigmapix_t = \sqrt{\Bar{\alpha}_t} \sigmapix_0 + \sqrt{1 - \Bar{\alpha}_t} \bepsilon, \quad \bepsilon \sim \mathcal{N}(\mathbf{0},\mathbf{I})
\end{align}
with $\Bar{\alpha}_t = \prod_{i=1}^T (1 -\beta_i)$. The goal of DDPMs is to reverse this diffusion process by learning parametrised transition densities $p_\theta(\sigmapix_{t-1}| \sigmapix_t)$. 
%Ho et al.\cite{ho2020denoising} propose to parametrise these transition densities as Gaussians with 
%\begin{align}
%    p(\sigmapix_{t-1}| \sigmapix_t; \theta) = \mathcal{N}(\sigmapix_{t-1}| \mu_\theta(\sigmapix_t, t), (1 - \Bar{\alpha}_t) I), 
%\end{align}
%where the mean is further parametrised as 
%\begin{align}
%    \mu_\theta(\sigmapix_t, t) = \frac{1}{\sqrt{1 - \beta_t}} \left( \sigmapix_t - \frac{\beta_t}{\sqrt{1 - \Bar{\alpha}_t}} \epsilon_\theta(\sigmapix_t; t) \right).
%\end{align}
Training a DDPM amounts to minimising the so-called $\bepsilon$-matching loss \cite{ho2020denoising}. %, which is equivalent to learning the score function \cite{vincent2011connection}.
This framework can be extended to conditional density estimation by including the measurements in the parametrised transition densities, i.e., $p_\theta(\sigmapix_{t-1}|\sigmapix_t, \delta \mathbf{U})$, using a conditional neural network $\bepsilon_\theta(\sigmapix_t, \delta \mathbf{U}; t)$ and minimise a conditional $\bepsilon$-matching loss
\begin{align}
    \min_\theta \E_{t \sim U(\{1, \dots, T\})} \E_{(\sigmapix_0, \delta \mathbf{U}) \sim p(\sigmapix_0, \delta \mathbf{U})}  \E_{\bepsilon \sim \mathcal{N}(\mathbf{0},\mathbf{I})}[ \| \bepsilon_\theta(\sigmapix_t, \delta \mathbf{U}; t) - \bepsilon \|_2^2],
\end{align}
with $\sigmapix_t$ as given in Eqn.~\eqref{eq:forward_onestep} and the expectation over $(\sigmapix_0, \delta \mathbf{U})$ is estimated using the simulated dataset. An overview of the network is given in Figure~\ref{fig:conddiffUNet}, where the input to the network is a concatenation of the linearised reconstructions, interpolated to the pixel grid, and the noisy image $\sigmapix_t$, together with the time step~$t$.

In \cite{ho2020denoising}, the authors make use of ancestral sampling to sample from the learned distribution. However, this requires to simulate the reverse process for all $T$ timesteps, resulting in a computationally expensive sampling method. To increase the sampling speed, we make use of the accelerated sampling scheme proposed in the DDIM framework \cite{song2020denoising}. Let $\tau$ be a subsequence of $\{1, \dots ,T\}$ of length $S$ with $\tau_1 = 1$ and $\tau_S = T$. The DDIM sampling, starting with $\sigmapix_{\tau_S} \sim \mathcal{N}(\mathbf{0},\mathbf{I})$, is given by 
\begin{align}
    \label{eq:ddim_sampling}
    \sigmapix_{\tau_{s-1}} = \sqrt{\Bar{\alpha}_{\tau_{s-1}}} \hat{\sigmapix}_0(\sigmapix_{\tau_s}, \delta \mathbf{U}) + \sqrt{1 - \Bar{\alpha}_t - \gamma_{\tau_s}^2}\bepsilon_\theta(\sigmapix_{\tau_s}, \delta \mathbf{U}, \tau_s) + \gamma_{\tau_s} \bepsilon,
\end{align}
with $\bepsilon \sim \mathcal{N}(\mathbf{0},\mathbf{I})$ and $\hat{\sigmapix}_0(\sigmapix_{\tau_s}, \delta \mathbf{U})$ as the Tweedie estimate \cite{efron2011tweedie}, defined by
\begin{align}
    \E[\sigmapix_0 | \sigmapix_{t}, \delta \mathbf{U}] \approx \hat{\sigmapix}_0(\sigmapix_{t}, \delta \mathbf{U}) = \frac{1}{\sqrt{\Bar{\alpha}_t}} \left( \sigmapix_t - \sqrt{1 - \Bar{\alpha}_t} \bepsilon_\theta(\sigmapix_t, \delta \mathbf{U}; t) \right).
\end{align}
The stochasticity parameter $\gamma_t$ in Eqn.~\eqref{eq:ddim_sampling} is chosen as 
\begin{align}
    \gamma_{\tau_s} = \eta \sqrt{(1 - \Bar{\alpha}_{\tau_{s-1}}) / ( 1 - \Bar{\alpha}_{\tau_s})} \sqrt{1 - \Bar{\alpha}_{\tau_s}/\Bar{\alpha}_{\tau_{s-1}}},
\end{align}
with a tunable hyperparameter $\eta$, see \cite{song2020denoising}.

In our implementation, we do not directly feed $\delta \mathbf{U}$ into the epsilon model $\bepsilon_\theta$, but rather make use of the initial reconstructions introduced in Section~\ref{sec:initial_reconstruction}. Thus, our model is of the form $\bepsilon_\theta(\sigmapix_t, \boldsymbol{\DeltaSigmaHat}, t)$ where $\boldsymbol{\DeltaSigmaHat}$ denotes the set of five interpolated linearised reconstruction in Eqn.~\eqref{eq:initial_reco}. In this way, we do not approximate the true posterior $p^\text{post}(\sigmapix| \delta \mathbf{U})$, but rather a conditional distribution~$p(\sigmapix |\boldsymbol{\DeltaSigmaHat})$.

As the goal was to produce a segmentation and not a reconstruction, we do not represent $\sigmapix$ using conductivity values, but rather as an image with values in $[0, 2]$. 
% \Simon{Maybe say "labels" or "integer values" rather than "values" to make it clear it is not a real number}. 
The segmentation is then obtained by rounding the reconstruction $\sigmapix$ to the nearest $\{0,1,2\}$ integer, where $0$ represents the background class, $1$ the resistive and $2$ the conductive inclusion class. 
For the final segmentation, we draw $J$ samples $\sigmapix^{(j)}$ using DDIM Eqn.~\eqref{eq:ddim_sampling} and perform a pixel-wise majority voting, i.e.,
\begin{align}
    \hat{\sigmapix}_i = \argmax_{c=0,1,2} \# \{ \sigmapix_i^{(j)} | \sigmapix^{(j)}_i = c, j = 1, \dots, J \},
\end{align}
for all pixels $i=1, \dots,d$ and where $\#$ denotes the cardinality of the set. 

\section{Practical consideration}
In this Section, we cover practical considerations of our submission. First, we cover the generation of the training data, second, we give details about the computation of the linearised reconstruction and third, we discuss aspects of the neural network architecture.

% \si{\Omega}^{-1}
\subsection{Dataset}
\label{sec:dataset}
\begin{figure}
    \centering
    \includegraphics[width=0.666\textwidth]{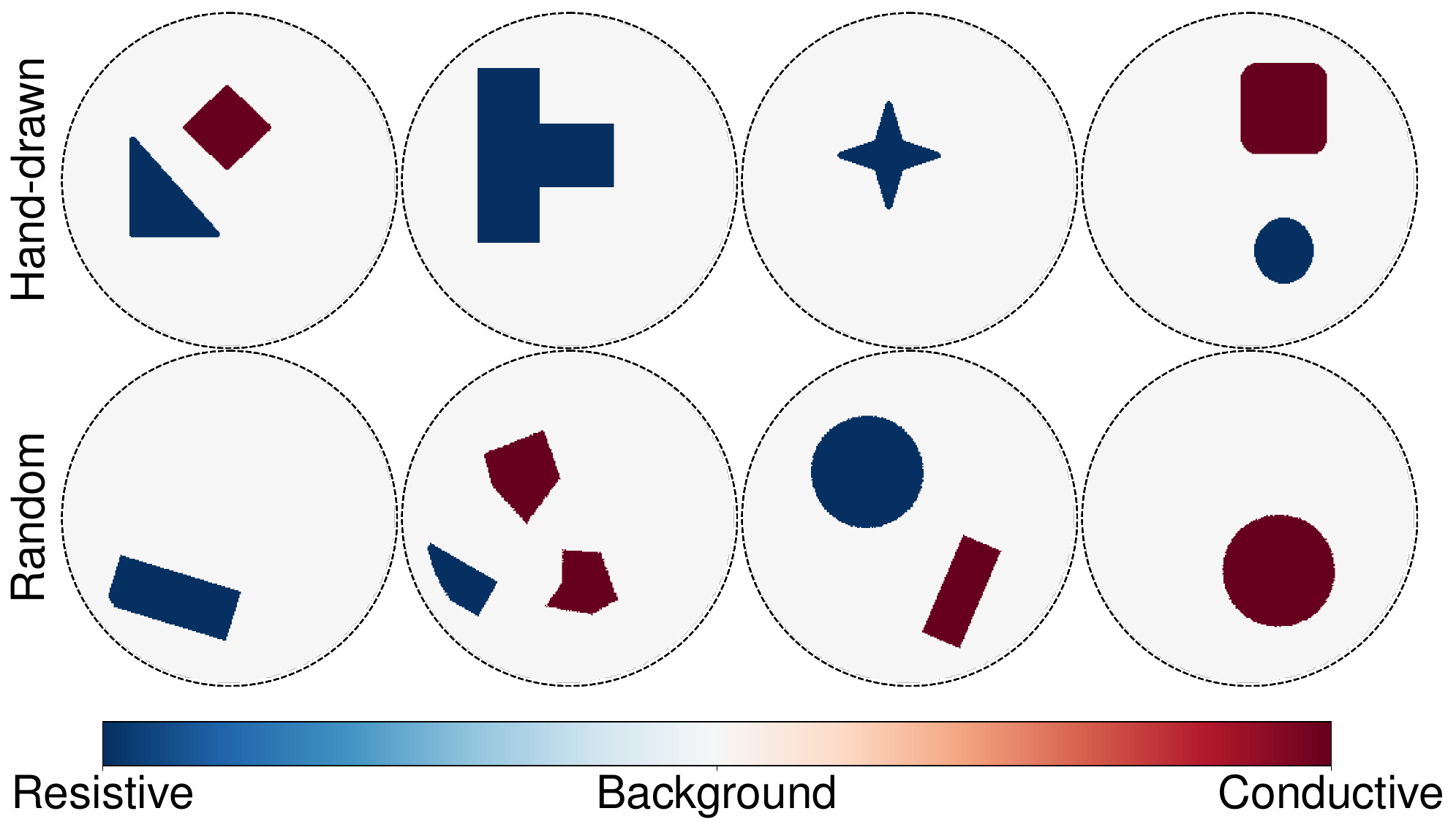} % 0.666
    \caption{Top: Handdrawn training phantoms. Bottom: Randomly generated training phantoms. For the visualisation, we only show the circular water tank. However, note that all models are trained using the square $256\times 256$ pixel images.}
    \label{fig:training_data}
\end{figure}

An important aspect of our submission is the creation of a simulated dataset suitable for training the different deep learning approaches. We started by generating random segmentation maps consisting of non-overlapping polygons, circles, rectangles and handdrawn objects on the~$256 \times 256$ pixel grid. Example phantoms are presented in Figure~\ref{fig:training_data}, where we only visualise the circular water tank. Each object was assigned to be either resistive or conductive. The areas outside of an object, and outside the water tank, were assigned the background class. Given this segmentation map, we simulate conductivity values for the objects. The conductivity of resistive objects was randomly chosen in $[0.025 ~\conductivity, 0.125 ~\conductivity]$ and the conductivity of conductive objects in $[5.0 ~\conductivity, 6.0 ~\conductivity]$. 
The background was assigned a constant conductivity value of $0.745 ~\conductivity$, which was computed using the reference measurements of the empty water tank via least squares fitting~\cite{vilhunen2002simultaneous}.  In the next step, the resulting phantoms were interpolated from the pixel grid to the piecewise constant finite element representation. The measurements were simulated using the forward operator specified in Section \ref{sec:forward_operator}. Gaussian noise was added with zero mean and covariance $\BSigma = \text{diag}(0.05 ~ \Uref + 0.01 \max(\Uref))$ according to the reference measurements of the empty water tank. For the simulation of the measurements a fixed contact impedance $z=\num{1e-6} \impedance$ was chosen for all $32$ electrodes\footnote{We also experimented with identifying the contact impedance using least squares fitting, but did not obtain good results.}. The number of training samples used per level is provided in Table~\ref{tab:num_training}. In total, we simulated more than $100$K data pairs. The lower number of training samples for level $6$ was due to technical problems in the simulation.

\begin{table}[ht]
\caption{Number of training samples used per level.}
\begin{tabular}{l|ccccccc}
Level            & 1 & 2 & 3 & 4 & 5 & 6 & 7 \\ \hline
Training samples & $16527$  & $16619$  & $16591$  & $16587$  & $16604$  & $12102$  & $16298$  
\end{tabular}
\label{tab:num_training}
\end{table}

\subsection{Initial Reconstruction}
\label{sec:practical_initReco}
Both the \PostP~ approach in Section~\ref{sec:post_unet} and the \CondD~ in Section~\ref{sec:cond_score} require an initial reconstruction as the input. We experimented with different classical reconstruction methods. Iterative reconstruction methods, e.g., $L1$-regularisation \cite{GEHRE20122126} or Gauß-Newton methods \cite{borsic2009vivo}, resulted in higher quality reconstructions compared to the linearised approach in Section~\ref{sec:initial_reconstruction}. However, as this initial reconstruction has to be computed for every example in the training set, i.e., for more than $100$K examples in the dataset we used, the computational expensive was a constraint. Thus, we decided against the computationally more expensive iterative methods and used the faster linearised reconstruction. However, even for the linearised reconstruction, simulation of the measurements and computation of the initial reconstruction took about a week.  

\begin{figure}[ht]
    \centering
    \includegraphics[width=0.3\linewidth]{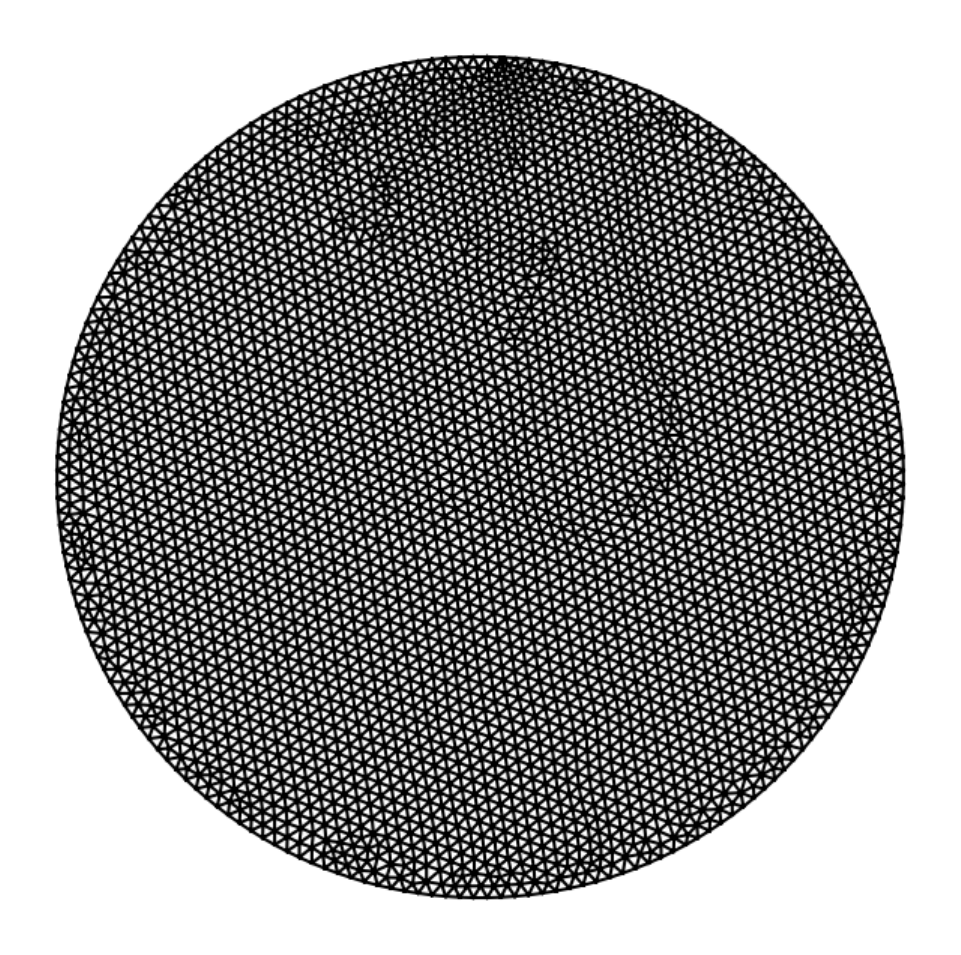}
    \hspace{1cm}
    \includegraphics[width=0.3\linewidth]{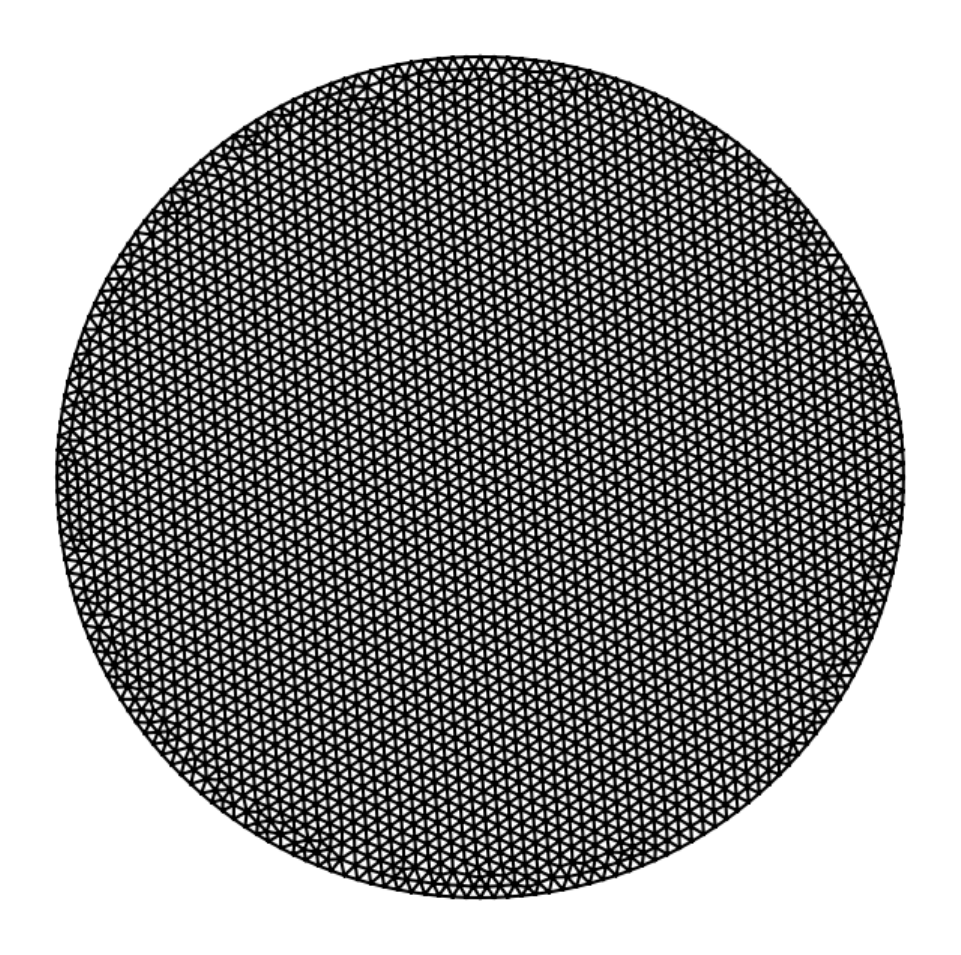}
    \caption{Left: The mesh provided by the organizers. Right: Our custom mesh for the forward operator.}
    \label{fig:mesh}
\end{figure}

The organisers provided a finite element implementation of the CEM. We decided to use our own implementation to more easily change the discrete function spaces and use a different mesh. A comparison of our mesh and the provided mesh is shown in Figure \ref{fig:mesh}. The provided mesh shows some small irregularities at the centre and top of the domain, which led to some differences in the forward solution and initial reconstructions. Instead, we use a uniform mesh with a mesh size of $0.005$, which was created with the software Gmsh \cite{gmsh2009}. Further, we set the boundary elements to cover the electrodes. 

%\TODO{Any important bits of the implementation? Choice of regularisers? Maybe talk a bit why we didnt use more sophisticated reconstructions methods ($L1$, Gauß Newton). $\to$ too slow.}

%\begin{figure}[ht]
%    \centering
%    \includegraphics[width=\textwidth]{images/training_data.pdf}
%    \caption{Left: Handdrawn training phantoms. Right: Randomly generated training phantoms}
%    \label{fig:training_data}
%\end{figure}

\subsection{Neural Network Architecture}
 For this the scalar value (level/timestep) is embedded into the architecture to reweight residual blocks depending on the scalar value, more effectively influencing the models output.

We use a minimally adapted guided diffusion model proposed by \cite{dhariwal2021diffusion}. The architecture consists of a U-Net \cite{ronneberger2015u} with attention blocks and time embedding. The time embedding was adapted for \PostP~ and \FullyC~ to allow the network to incorporate level information, meaning the training data across all levels can be used during training. The time or level information is introduced to the network by adaptive group normalisation layers~\cite{dhariwal2021diffusion}. Each group normalisation layer~\cite{wu2018group} in the U-Net is replaced with 
\begin{align}
    \text{AdaGroupNorm}(h, z) = z_s \text{GroupNorm}(h) + z_t,
\end{align}
where $h$ is the intermediate feature and $z = (z_s, z_t)$ is the output of a neural network taking the time or level information as an input. With our choice of hyperparameters, e.g., number of layers, channels, etc., the total number of trainable parameters is $31$M. % 31049731

The number of input and output channels of the U-Net varied between the approaches. \PostP~ and \CondD~ had five input channels corresponding to the five interpolated linearised reconstructions $\boldsymbol{\DeltaSigmaHat}$, whereas \FullyC~ had a single channel input for the interpolated learned reconstruction $\mathcal{S}(W\delta\mathbf{U})$. For the learned reconstructors a CCE loss was used that required the three class probabilities, thus three output channels were used. For \CondD~ the $\bepsilon$-matching loss was used, requiring a single channel output. Due to differences in input/output channels each U-Net backbone did not have an equal number of parameters, albeit the difference was negligible. For \FullyC~ the linear layer $\mathcal{W}$ required 10M parameters; this is a significant increase in learnable parameters as compared to the other approaches.

\section{Results and Discussion}
\begin{figure}[h!]
    \centering
    \includegraphics[width=0.666\textwidth]{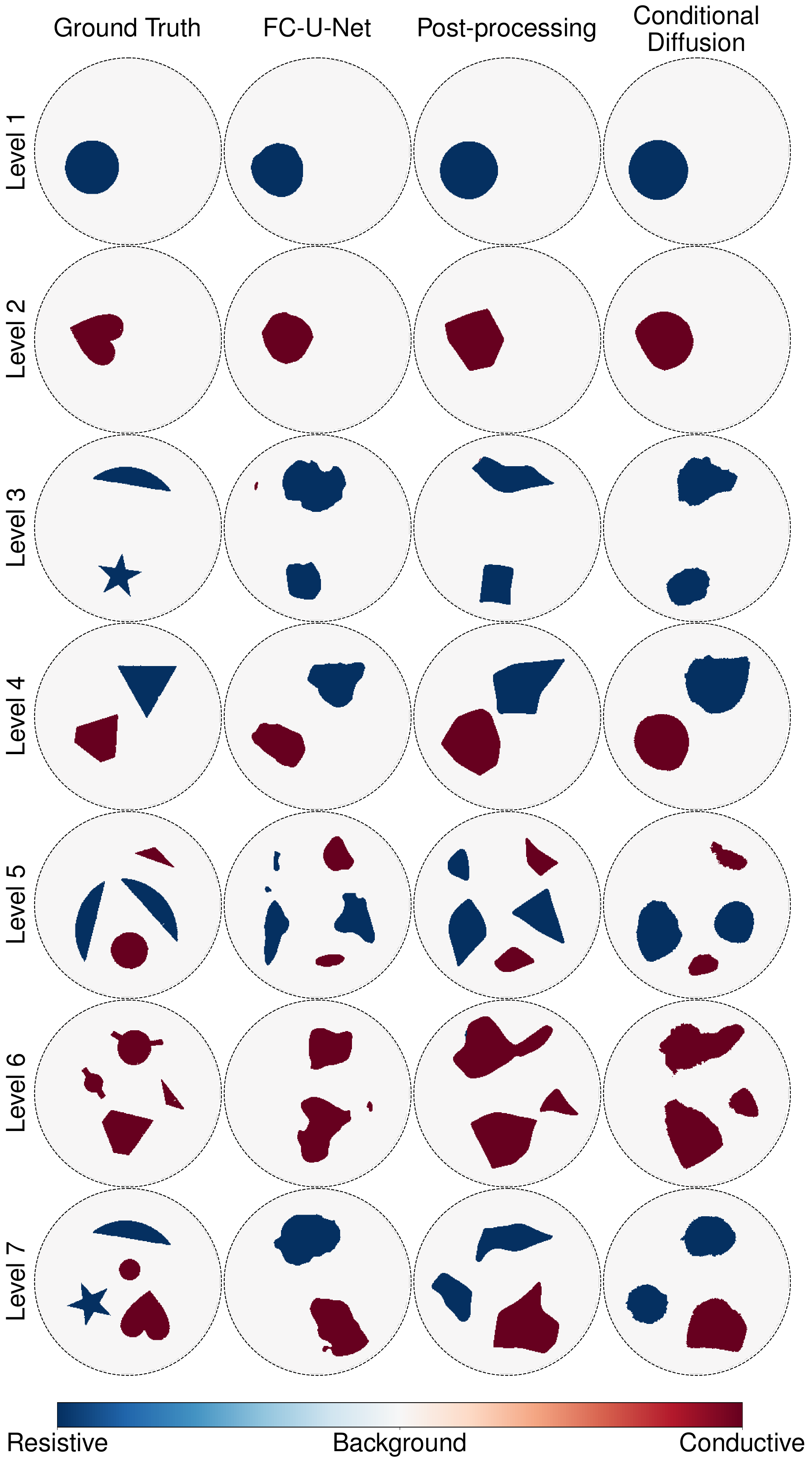} % 0.666
    \caption{Segmentation of the three methods for sample $A$ of level $1$ to $7$.}
\label{fig:reco_sample_A}
\end{figure}

\begin{figure}[h!]
    \centering
    \includegraphics[width=0.666\textwidth]{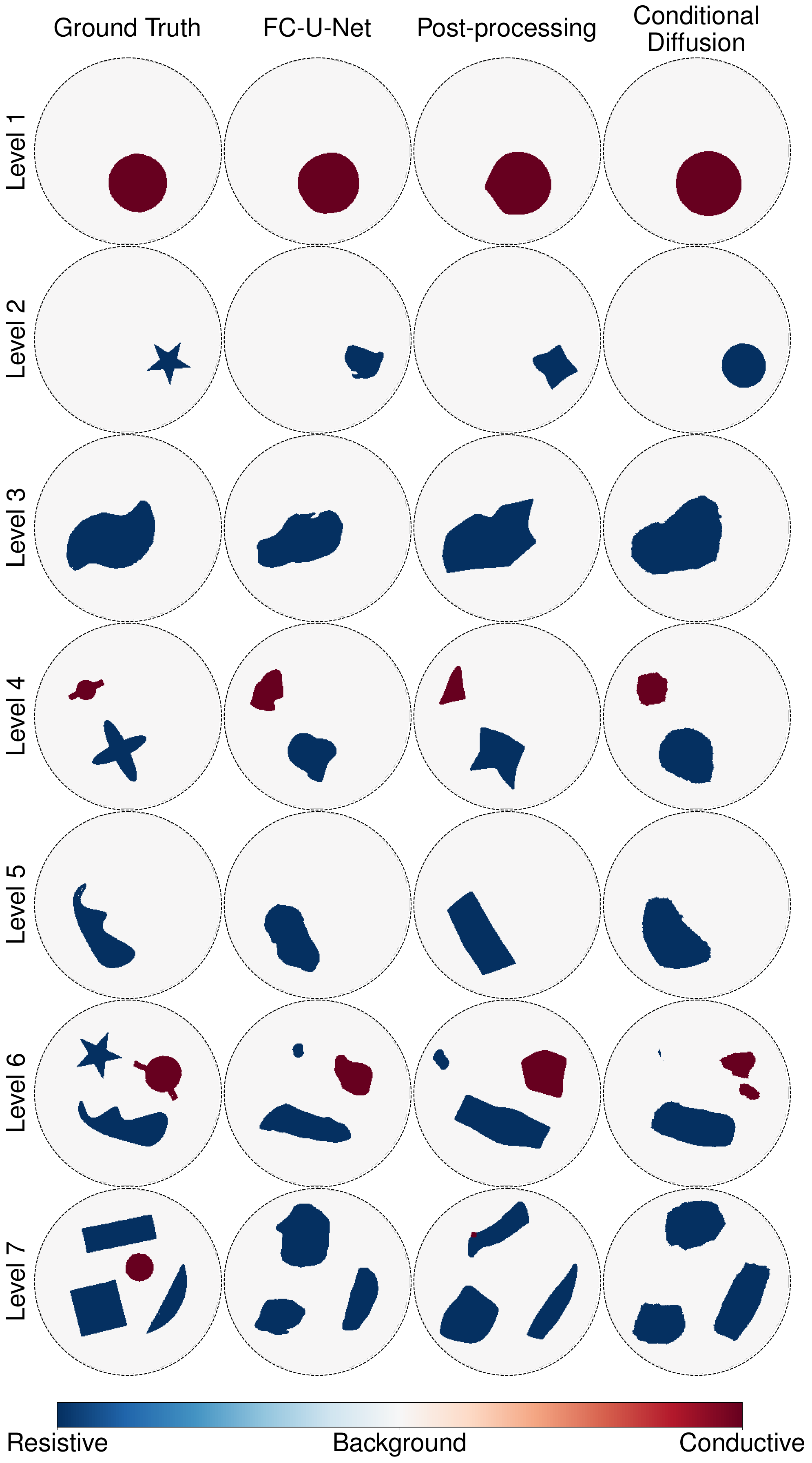}
    \caption{Segmentation of the three methods for sample $B$ of level $1$ to $7$.}
    \label{fig:reco_sample_B}
\end{figure}

\begin{figure}[h!]
    \centering
    \includegraphics[width=0.666\textwidth]{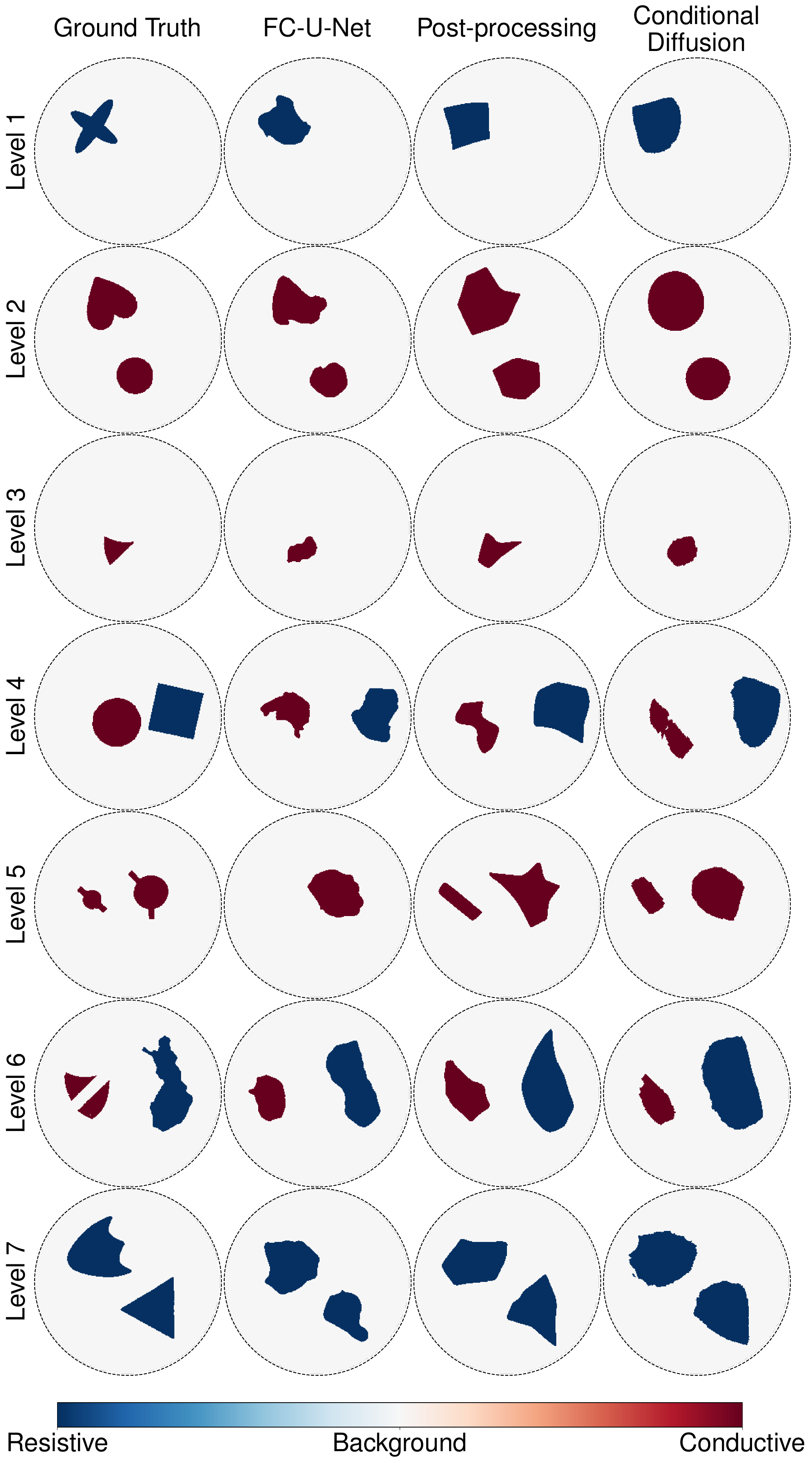}
    \caption{Segmentation of the three methods for sample $C$ of level $1$ to $7$.}
    \label{fig:reco_sample_C}
\end{figure}

In this section we present the final challenge results for our three approaches. Quantitative scores are computed using structural similarity index measure (SSIM) \cite{wang2004image} individually on binary maps of conductive and resistive inclusions. This was averaged to give a per-sample score. This per-sample score was summed across all samples of a level to give a level score, and then summed over all levels to give the overall score of the methods\footnote{Three phantoms were evaluated per level resulting in a maximum score of $21$.}. For each challenge level, three different phantoms (A,B,C) were evaluated. Visual results are presented in Figure~\ref{fig:reco_sample_A}, Figure~\ref{fig:reco_sample_B} and Figure~\ref{fig:reco_sample_C}. In most reconstructions, the number of objects, positions and rough shape are correctly identified. Exceptions are cases where a small conductive object was placed in the middle of the water tank and surrounded by other objects, see for example level $7$ in Figure~\ref{fig:reco_sample_A} or level $7$ in Figure~\ref{fig:reco_sample_B}. Further, the reconstruction of objects on the upper left side of the water tank is often worse as the measurements of this part boundary are removed for higher levels. See for example level $7$ in Figure~\ref{fig:reco_sample_B}, where the shape of the rectangle at the top of the water tank can not be recovered. 

Quantitative results are presented in Table~\ref{tab:full_results}. Besides our submission, we also present the results of the second and third best performing team. With a final score of $15.24$ the \PostP~ approach was the best performing method in \KTC. However, the \FullyC~ was able to outperform this approach at levels $2,4,5$ and $6$. The second place with a score of $12.75$ was achieved by a team of the federal University of ABC and the third place was achieved by a team from DTU with a score of $12.45$. On level $4$ the second place even achieved a higher score than the \PostP. Both our \FullyC~ approach, with a score of $15.13$, and our \CondD, with a score of $14.60$, would have won the challenge. 

\PostP~ and the \FullyC~ perform similarly, while worse performance can be observed with \CondD. For the \CondD~ approach, a separate neural network was trained for each challenge level. Thus, the network for each level was only trained using a subset of all available phantoms and measurements that were simulated. Whereas both \PostP~ and \FullyC~ approaches utilised the training examples across all levels. The learned reconstructor approaches utilised CCE loss specific to segmentation tasks, whereas \CondD~ used a $\bepsilon$-matching which is required for DDPM. Rather than using a single sample, for \CondD~ $J$ conditional samples were drawn and the segmentation was determine via majority voting, this could be extended to obtain a notion of uncertainty.

The \PostP~ and \CondD~ approaches both took a set of five linearised reconstructions as input. Through using a set of reconstructions with different regularisation strengths we attempt to obtain a more robust segmentation as the best regularisation strength is not known. In a similar sense, a set of reconstructions could be learned with the \FullyC~ but would require significant increase in the number of learnable parameters. %All methods used linearised reconstruction, whether learned or classical, \Simon{Something missing in this sentence; it doesn't quite make sense} it would be of interest to use the discrete non-linear forward operator for reconstruction; we leave this as further work.

\begin{table}[]
\caption{Quantitative comparison of our three submissions via structural similarity index measure (SSIM). These are official challenge results, rounded to the nearest hundredth. The second place was achieved by Team ABC from the Federal University of ABC, Brasil. The third place was achieved by Team DTU from Technical University of Denmark. SSIM is averaged for a given sample between conductive and resistive inclusions. At each level the SSIM is summed across the three samples, and the overall sum for a method is summed across all samples and levels.}
\label{tab:full_results}
\begin{tabular}{l|ccccccc|c}
Level    & 1 & 2 & 3 & 4 & 5 & 6 & 7 & Sum \\ \midrule
\FullyC & $2.72$  & {\cellcolor[gray]{.9}$2.64$} & $2.31$  & {\cellcolor[gray]{.9}$1.80$}  & {\cellcolor[gray]{.9}$2.06$}   & {\cellcolor[gray]{.9}$2.07$}   & $1.53$   & $15.13$          \\
\PostP & {\cellcolor[gray]{.9}$2.76$}  & $2.56$  & {\cellcolor[gray]{.9}$2.54$}   & $1.71$  & $2.06$ & $1.92$   & {\cellcolor[gray]{.9}$1.69$}  & {\cellcolor[gray]{.9}$15.24$}  \\
\CondD & $2.67$ & $2.49$ & $2.47$ & $1.61$ & $1.94$ & $1.76$ & $1.65$ & $14.60$ \\ \midrule 
Team ABC & $2.75$ & $2.37$ & $2.07$ & $1.74$ & $1.08$ & $1.53$ & $1.22$ & $12.75$ \\ 
Team DTU & $2.28$ & $2.3$ & $1.87$ & $1.55$ & $1.34$ & $1.44$ & $1.60$ & $12.45$ \\ 
\end{tabular}
\end{table}

\section{Conclusion}
The \KTC~ challenge provided an opportunity to evaluate state-of-the-art methods on the problem of reconstructing segmentation maps from EIT measurements. Our winning submissions utilised deep learning, with two learned reconstructor methods (\FullyC~ and \PostP), as well as a \CondD~ generative method. The choice of network architecture and dataset are vitally important for deep learning approaches; requiring knowledge of the problem whilst being practical. In this work all submissions utilised the same training dataset and back-bone network structure; allowing for comparison between methods. Both \FullyC~ and \PostP~ provided similar results, whereas \CondD~ performed less well. The learned reconstructors were trained across all levels (utilising level-conditioning), whereas the individual \CondD~ networks were trained individually for each level, effectively reducing the training dataset size. The \FullyC~ required an additional fine-tuning phase on the provided real measurements and phantom, this was not needed for the \PostP~ network which only used simulated measurements and phantoms. The \PostP~ and \CondD~ methods took a set of five Tikhonov-regularised initial reconstructions as input, while the \FullyC method used linear layer to map from measurement to a single image. Out of the three methods submitted the \PostP~method gave the best performance. This suggests that a post-processing approach trained on a high-quality simulated data set can generalise to real data more easily than a fully learned method. Further work is necessary to fully evaluate the generalisation capabilities of different data-driven approaches. 

\section*{Acknowledgments}
Alexander Denker acknowledges the support by the Deutsche Forschungsgemeinschaft (DFG) within the framework of GRK 2224/1 ``$\pi^3$: Parameter Identification -- Analysis, Algorithms, Applications''. Tom Freudenberg acknowledges funding by the DFG – Project number 281474342/GRK2224/2. Imraj Singh was supported by the EPSRC-funded UCL Centre for Doctoral Training in Intelligent, Integrated Imaging in Healthcare (i4health) (EP/S021930/1) and the Department of Health’s NIHR-funded Biomedical Research Centre at University College London Hospitals. \v{Z}eljko Kereta was supported by the UK EPSRC grant EP/X010740/1. Simon Arridge is supported by the UK EPSRC EP/V026259/1. Peter Maass is supported by BMBF-project ``\#MOIN - MuKIDERM''. T. Kluth acknowledges support from the KIWi project funded by the German Federal Ministry of Education and Research (Bundesministerium für Bildung und Forschung, BMBF, project number 05M22LBA).

%%%%%%%%%%%%%%%%%%%%%%%%%%%%%%%%%%%%%%%%%%%%%%%%%%%%%%
%          7. REFERENCES SECTION
%%%%%%%%%%%%%%%%%%%%%%%%%%%%%%%%%%%%%%%%%%%%%%%%%%%%%%

%       READ THIS SECTION CAREFULLY

% Each of the references below MUST be cited in your article above. Do not include references that are not cited in your article.

% Follow the examples below carefully. We strongly suggest that you copy and paste your reference information directly into our examples.

% List all references in alphabetical order according to the first author's last name.

% Verify each URL works correctly and can be accessed properly. Your URL links should be to reputable websites. The command line for a website link begins with: \url{ }

% Do not add MR or DOI numbers to your references. AIMS production staff will add this information.

% Using BibTex is not recommended but can be handled.

\medskip
% The information below will be filled in by AIMS production staff.
Received xxxx 20xx; revised xxxx 20xx; early access xxxx 20xx.
\medskip

\end{document}